\documentclass[sigconf]{acmart}
\renewcommand\footnotetextcopyrightpermission[1]{} 
\input{preamble.sty}
\newcommand{\setOfReals}{\mathbb{R}}
\newcommand{\setOfNaturals}{\mathbb{N}}

\newcommand{\setOfPositiveReals}{\setOfReals_{+}}

\newcommand{\brac}[1]{\left({#1}\right)}
\newcommand{\sbrac}[1]{\left[{#1}\right]}
\newcommand{\cbrac}[1]{\left\{{#1}\right\}}

\newcommand{\expect}[1]{\mathbb{E}\left[{#1}\right]}

\newcommand{\UN}[1]{{\mathcal{U}}^{(N)}_{j}}

\newcommand{\norm}[1]{\|{#1}\|}
\newcommand{\abs}[1]{\left \vert {#1} \right \vert}
\newcommand{\mb}[1]{\mathbb{#1}}

\newcommand{\mc}[1]{\mathcal{#1}}

\newcommand{\floor}[1]{\left\lfloor{#1}\right\rfloor}

\DeclareMathOperator*{\argmin}{arg\,min}
\DeclareMathOperator*{\argmax}{arg\,max}

\makeatletter
\newcommand{\distas}[1]{\mathbin{\overset{#1}{\kern\z@\sim}}}%
\newsavebox{\mybox}\newsavebox{\mysim}
\newcommand{\distras}[1]{%
  \savebox{\mybox}{\hbox{\kern3pt$\scriptstyle#1$\kern3pt}}%
  \savebox{\mysim}{\hbox{$\sim$}}%
  \mathbin{\overset{#1}{\kern\z@\resizebox{\wd\mybox}{\ht\mysim}{$\sim$}}}%
}
\makeatother

\newboolean{todo}
\setboolean{todo}{true} 
\newcommand{\remarkInternal}[4]{\ifthenelse{\boolean{todo}}{\todo[inline, color=#2, caption={2do}, #3]{\begin{minipage}{\textwidth-4pt}\emph{Remark #1:}\\#4\end{minipage}}}{}}

\AtBeginDocument{%
  \providecommand\BibTeX{{%
    \normalfont B\kern-0.5em{\scshape i\kern-0.25em b}\kern-0.8em\TeX}}}

\setcopyright{acmcopyright}
\copyrightyear{2018}
\acmYear{2018}
\acmDOI{10.1145/1122445.1122456}

\begin{document}


\title{On the Throughput Optimization in Large-Scale Batch-Processing Systems
}



\author{Sounak Kar}
\affiliation{%
  \institution{TU Darmstadt}
  \city{Darmstadt}
  \country{Germany}
}

\author{Robin Rehrmann}
\affiliation{%
  \institution{TU Dresden}
  \city{Dresden}
  \country{Germany}
}

\author{Arpan Mukhopadhyay}
\affiliation{%
  \institution{University of Warwick}
  \city{Coventry}
  \country{United Kingdom}
}

\author{Bastian Alt}
\affiliation{%
  \institution{TU Darmstadt}
  \city{Darmstadt}
  \country{Germany}
}

\author{Florin Ciucu}
\affiliation{%
  \institution{University of Warwick}
  \city{Coventry}
  \country{United Kingdom}
}

\author{Heinz Koeppl}
\affiliation{%
  \institution{TU Darmstadt}
  \city{Darmstadt}
  \country{Germany}
}

\author{Carsten Binnig}
\affiliation{%
  \institution{TU Darmstadt}
  \city{Darmstadt}
  \country{Germany}
}

\author{Amr Rizk}
\affiliation{%
  \institution{Universit{\"a}t Ulm}
  \city{Ulm}
  \country{Germany}
}


\begin{abstract}
We analyze a data-processing system with $n$ clients producing jobs which are processed in \textit{batches} by $m$ parallel servers; the system throughput critically depends on the batch size and a corresponding sub-additive speedup function. In practice, throughput optimization relies on numerical searches for the optimal batch size, a process that can take up to multiple days in existing commercial systems. In this paper, we model the system in terms of a closed queueing network; a standard Markovian analysis yields the optimal throughput in $\omega\left(n^4\right)$ time. Our main contribution is a mean-field model of the system for the regime where the system size is large. We show that the mean-field model has a unique, globally attractive stationary point which can be found in closed form and which characterizes the asymptotic throughput of the system as a function of the batch size. Using this expression we find the \textit{asymptotically} optimal throughput in $O(1)$ time. Numerical settings from a large commercial system reveal that this asymptotic optimum is accurate in practical finite regimes.    
\end{abstract}

\maketitle

\section{Introduction}
A key technique to cutback overhead in data-processing systems is \textit{service batching}, i.e., collecting the inputs to form batches that are then processed as one entity.
The rationale lies in the overhead amortization with increasing the batch size.
A prominent example highlighting the benefits of service batching is a Linux-based system in which the network-card throughput can be substantially increased by batching data packets~\cite{linux-packetbatching}.
Similar improvements hold in software-defined networks by passing switching rule updates in batches from controllers to network switches~\cite{Wen-ICDCS16}.
In this work, we analyze the benefits of service batching in the context of large-scale data-processing systems, and in particular of a large commercial database system.

\begin{figure}[t]
	\centering
	\includegraphics[width=0.99\linewidth]{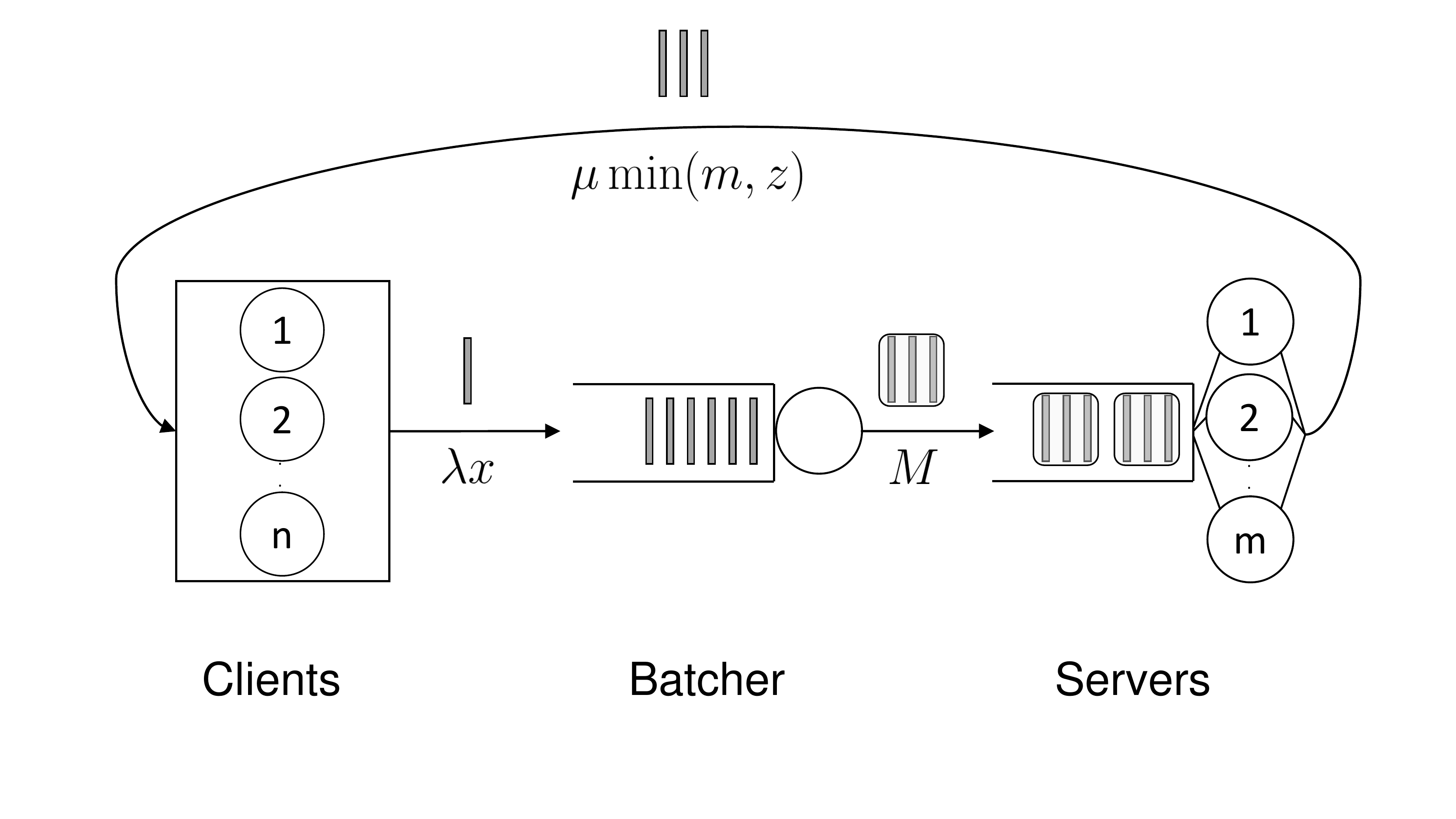}
    \caption{A closed queueing system with $n$ clients and $m$ servers. Clients are either active or inactive and produce jobs at rate $\lambda x$ when $x$ of them are \emph{active}. The batcher produces batches of size $k$ at rate $M\floor{y/k}$ when there are $y$ available jobs. The service station consists of a single queue and $m$ parallel servers, each having a service rate $\mu$; the overall \emph{batch} service rate is $\mu\min(m,z)$ when $z$ batches are available. 
    }
    \label{fig:closedSystemDesc}
    \vspace{-10pt}
\end{figure}

We consider a closed system in which $n$ clients generate jobs to be processed by $m$ parallel servers. Each client alternates between being in either an \emph{active} or an \emph{inactive} state; in the former it produces a job and in the latter it awaits the job to be fully processed. We note that each client can have at most one job in the system, i.e., a client produces a new job no sooner than its previous one finished execution. The servers process jobs in batches of size $k$, i.e., once $k$ clients produce $k$ jobs these are sent for batch processing -- and may have to wait in a central queue if all servers are busy; see Fig.~\ref{fig:closedSystemDesc}\footnote{All times are exponentially distributed with the rates $\lambda$, $M$, and $\mu$, the last two depending on the batch size $k$; we will show that this technically convenient assumption is valid by fitting our model's parameters from a real-world system.}. This model is representative for some real-world data-processing systems such as databases employing Multi Query Optimization  \cite{Thomson:2012:CFD:2213836.2213838:Calvin,rehrmann2018oltpshare,Sellis:1988:MO:42201.42203:MQO}. 

Besides a model with a single job type, we also consider a generalized model with  two job types. A typical example would be \textit{read} and \textit{write} jobs in a database system; such jobs not only have different average processing times but some are prioritized over the others, e.g., the \textit{write} jobs have non-preemptive priority over the \textit{read} jobs for consistency reasons.

Classical approaches to queueing systems with batch arrivals and batch service disciplines have been intensively studied, e.g., in \cite{bailey1954queueing,deb1978optimal,chaudhry1983first,Bolch:2005} and the references therein. Most of these studies were either mainly concerned with open queueing systems or focused on different properties of interest such as the product form; for a more thorough discussion see Sect.~\ref{sec:related_work}. To the best of our knowledge the closed queueing system from Fig.~\ref{fig:closedSystemDesc} is new, i.e., it does not fit existing models.

The main contribution of this paper consists in the throughput optimization in a closed batching system characteristic to a large production system; this involves finding the optimal batch sizes. We first provide the exact analysis by solving for the balance equations in a Markov model, an approach requiring at least $\omega\left(n^4\right)$ computational time. We also provide the corresponding mean-field models which yield exact results in an asymptotic regime whereby both $n$ and $m$ are proportionally scaled. This second approach yields the optimal (asymptotic) throughput in $O(1)$, which is particularly appealing given that existing  empirical approaches rely on extensive numerical searches for the optimal batch sizes, a process which typically runs in the order of days\footnote{According to personal communications with engineers from a large commercial database system}.

To find the asymptotically optimal batch size, we first prove that the dynamics of the system converges to a deterministic mean-field limit as $n,m \to \infty$. We then find a closed form solution of the stationary point of the mean-field and prove that it is globally attractive. Using the stationary point of the mean-field we characterize the throughput of the system as a function of the batch size. This finally leads to a simple optimization problem which can be solved either in closed form or numerically in constant time to find the asymptotically optimal batch size. 

Recently, mean-field techniques have been used successfully in various models of large scale service systems, such as web server farms \cite{arpan2016tcns}, cloud data centers \cite{xie2015sigmetrics}, and caching systems \cite{gast2015sigmmetrics}, where an exact solution of the stationary distribution is computationally infeasible due to the large size of the state space. In such systems, the key idea is to approximate the Markovian dynamics of the system by a deterministic dynamical system, called {\em the mean-field limit}, typically described by a system of ordinary differential equations (ODEs). Such an approximation is exact in the limit as the system becomes large. The stationary behaviour of the limiting system can be described by the stationary point of the mean-field which can either be found in closed form or computed in  constant time. The key challenge is to prove the uniqueness and existence of the stationary point and the fact that all possible trajectories of the mean-field limit converges to this unique stationary point (global attraction) \cite{budhiraja2015ejp,benny2019sigm}.       

To demonstrate the practical relevance of our results we analyze a large commercial database system. In such a system a job refers to a query, e.g., an SQL string, which can execute \textit{read} or \textit{write} operations. A client can only send a new query once the previous query has been processed, i.e., each client can have at most one outstanding query at any time. Job/query batching involves merging multiple similar queries into a new SQL string, whose execution time depends on many factors such as the operations' types. Moreover, the shared overhead amongst the individual queries lends itself to a certain speedup in the batch execution time which was empirically shown to be around a factor of $2$ in~\cite{rehrmann2018oltpshare}; the speedup is generally a function of both the number of batched jobs $k$ and the jobs' types, e.g. \textit{read} or \textit{write}. 

The remainder of the paper is structured as follows. We first discuss related work and then describe the queueing model and the optimization formulation. In Sect.~\ref{sec:meanField} we provide the mean-field model and the corresponding asymptotic result. In Sect.~\ref{sec:modelMultJobTypeBig} we provide the generalized model for the two types of jobs case, and then present numerical and experimental evaluation results for the optimal batch sizing approach in Sect.~\ref{sec:numerical}. Lastly we conclude the paper in Sect. ~\ref{sec:conclusion}. 
\section{Related Work}
\label{sec:related_work}
We overview some open and closed queueing systems with batching, and practical approaches to batching in database systems.

In the open queueing systems literature, one of the earliest examples of batching is \cite{bailey1954queueing} which derives the expected value of the steady state queue length and waiting time assuming exponential inter-arrival and Chi-squared service time. In \cite{deb1973optimal}, the authors consider a queueing system with Poisson arrivals and general batch service time, independent of the batch size; both the execution time and batch size can be dynamically controlled subject to real-world constraints on the maximum possible batch size. If a batch is forwarded to the server only at the points when the server is free, or there is an arrival or departure, it is shown that it is optimal to serve all jobs in a batch only when the queue length exceeds a certain threshold. Batching in the context of running a shuttle service between two end points has been considered in \cite{deb1978optimal}, which provides an optimal batching policy for minimizing the expected total discounted cost over an infinite horizon. Here it is assumed that the customers arrive according to independent Poisson processes. The authors in \cite{berg1998optimal} consider a discrete time system with incoming jobs having a strict delay guarantee. Given a certain form of serving cost which incentivizes batching and arrival distribution, the authors lay down a strategy that minimizes the expected long term cost per unit time. Further, in \cite{glazer1987equilibrium}, a queueing system with bulk service at scheduled time points has been considered where the customers can pick their arrival time to minimize the waiting time. Under some given conditions, the authors show that it is optimal to arrive just the moment before a service starts.

In turn, a key objective in the closed queueing systems literature was proving the product form property of the steady state queues' distribution. Gordon and Newell \cite{gordon1967closed} considered a closed network with multiple service stages and a set of probabilities governing the routing among these stages and showed the product form property under the assumption of exponential service times. In the seminal work on BCMP networks \cite{baskett1975open}, the authors considered the more general case of open, closed, and mixed networks, and also multiple job classes. Inspired by the functioning of central processors, data channels, terminals among others, sufficient conditions have been provided for each of these cases for the network to have a product form equilibrium distribution. Further, in \cite{chandy1977product}, the authors generalized the idea of local balance to station balance that explains the conditions for a network with non-exponential service times to have a product form. These findings were further extended under a more general set-up in \cite{chandy1983characterization}, which investigated the existence of product form equilibrium distribution under certain restrictions on the service discipline which can however be class dependent. The existence of product form in closed queueing networks with service batching was investigated in \cite{henderson1990closed}, which derives conditions for the existence of product form distribution in a discrete-time setting with state-independent routing, allowing multiple events to occur in a single time slot. The results were further extended to a continuous-time setting allowing for batch arrivals in \cite{henderson1990product}. For the particular closed queueing network with service batching from this paper, it is certainly of interest to determine whether the product form property applies. 
However, aforementioned works do not apply to our problem as the conditional routing probabilities of jobs/batches in our case is state-dependent due the FCFS nature of service. Further, even if we approximate FCFS order by random service order, we cannot \emph{directly} compute the system throughput from these works as they lack a method to derive the normalizing constant for the corresponding product form.

In the context of batching in databases, one of the earliest and influential work is \cite{dewitt1984groupcommit} whereby transactions are executed as sequence of jobs and batches of jobs access the same log page. Once that page is full, the log is flushed and the batch is executed, thus decreasing the I/O. Naturally, the batch size is fixed to the page size; in turn, in our work, we allow for flexible batch sizes in relation to the number of clients and specifically focus on optimizing throughput rather than I/O reduction. In comparison, \textit{SharedDB} \cite{giannikis2012shareddb} executes all incoming jobs as a big batch. Jobs that enter the system, while a batch is executed, are queued and batched, once the previous batch finished execution. In contrast to our work, SharedDB executes batches of different sizes sequentially and does not classify job types or consider job sizes.
A similar work to SharedDB is \textit{BatchDB} \cite{makreshanski2017batchdb} in which incoming analytic jobs are batched where the execution is interleaved with writing jobs, as they occur. Alike SharedDB, BatchDB does not classify their jobs or focus on the size of batches in relation to clients.
The closest system to our work is \textit{OLTPShare} \cite{rehrmann2018oltpshare}, where the authors use a fixed time interval to collect incoming jobs into batches. 
In contrast, our approach of using a count-based batching (i.e., each batch has exactly $k$ jobs) has the practical benefit of utilizing cached batch queries. These batch jobs are compiled SQL strings that have been requested before. 
Using the interval approach results in batches of various sizes diminishing the efficiency of caching previously seen batch requests. 
\section{Queueing Model and Optimization Goal}\label{sec:modelSingleJobType}
We consider a closed queueing system where jobs are routed along three stations: job producer, job batcher, and service station. The producer station has $n$ clients, each being assigned a token enabling them to submit a new job/query\footnote{We use the terms \textit{job} and \textit{query} interchangeably.}. Upon submission, the token is revoked and the query is passed to the job batcher which creates a merged query at rate $M(k)$, once $k$ queries become available to form a batch of size $k$. Each batch is forwarded to the service station consisting of $m$ serving units, or servers, processing batches in a FCFS order at rate $\mu(k)$, i.e., the number of \emph{batches} served per unit time. Further, the merged query is compiled, executed, and the result is split and sent back to the respective clients. Along with receiving a result, each client also receives its token back and becomes ready to submit a new query. We note that the rate at which a new query is submitted to the batching station depends on the number of \emph{active} clients, i.e., clients with a token, rather than the total number of clients. It is also important to observe that the total number jobs in the system is the same as the number of clients $n$. For a schematic representation of the system recall Fig.~\ref{fig:closedSystemDesc}. 

A key observation is that the additional time spent on batching is compensated by the reduction in the total execution time of the jobs, owing to the amortization of associated operational overhead characteristic to jobs of the same type. The gain from batching usually grows when increasing the batch size, an effect which is commonly referred to as \textit{speedup}. However, increasing the batch size beyond a certain threshold can lead to an excessive idling of the available servers. This is due to the fact that batch formation takes longer and also the number of batches in the system can become less than the number of servers. In other words, higher speedups can idle more servers, which raises an interesting performance tradeoff. Our objective is to find the  optimal batch size $k^*$ maximizing the system's throughput, i.e., the number of jobs served at the service station per unit time. To this end, we will first model the closed queueing system as a continuous time Markov chain (CTMC) and find its steady state distribution.

We assume that the time for each client to produce a job is exponentially distributed with rate $\lambda$; 
denoting by $x$ the number of active clients (i.e., having a token), the producer station forwards a job to the batcher at rate $\lambda x$. Let us also denote by $y$ and $z$ as the number of jobs at the batcher and the number of batches at the server, respectively. The state of the system can thus be uniquely described by the triple $(x,y,zk)$ belonging to the state space $$\mc{S}=\cbrac{(x_1,x_2,x_3)\in \mb{Z}_+^3: x_1+x_2+x_3 = n, k|x_3}~.$$ 
Although $(x,y,zk)$ is determined by any two of its components, we retain the triple representation due to a more convenient visualisation. The state of the system clearly evolves as a continuous-time Markov chain and the rates at which the system jumps to another state from the state $(x,y,zk)$ are given by
\begin{align}
\label{eq:jumpRates}
    (x,y,zk) &\xrightarrow[\text{}]{\lambda x} (x-1,y+1,zk), \thinspace x>0 \nonumber \\
            &\xrightarrow[\text{}]{M(k)\floor{y/k}} (x,y-k,(z+1)k), \thinspace y \ge k \nonumber \\
            &\xrightarrow[\text{}]{\mu(k)\min(m,z)} (x+k,y,(z-1)k), \thinspace z>0~.
\end{align}

Informally, when the system is in state $(x,y,zk)$, either one job can move from the producer to the batcher at rate $\lambda x$ when there are $x$ active clients, \emph{or} $k$ jobs can move from the batcher to the server at rate $M(k)\floor{y/k}$, \emph{or} $k$ more clients become active (i.e., receive their tokens back) at rate $\min(m,z)\mu(k)$. The rates to all other states are zero. 

The system attains a steady state with the unique distribution $\pmb{\pi}_0$ given by the solution of the equation $\pmb{\pi} \cdot \mathbf{Q} = 0$. This is due to the fact that the chain is irreducible, whereas the finiteness of the state space guarantees positive recurrence; for a rigorous argument see Sect. ~\ref{sec:app-0} in the Appendix. Here, $\mathbf{Q}(r,s)$ denotes the jump rate from state $r$ to $s$ where $r$, $s$ are of the form $(x,y,zk)$, as specified in \eqref{eq:jumpRates}. Given the non-linear
state dependent rates, we can only obtain the solution $\pmb{\pi}_0$ numerically rather than in closed form.

Further, the steady state distribution  $\pmb{\pi}$ immediately lends itself to the steady state system throughput, i.e.,
\begin{align}\label{eq:steadyThroughput}
    \Theta(k) := \sum_{(x,y,zk) \in \mc{S}} \pmb{\pi}_0(x,y,zk) \thinspace k\mu(k)\min(m,z)~,  
\end{align}
which implicitly yields the optimal batch size
\begin{align}
\label{optFormulation}
    k^* := \argmax_{k \in \mc{K}} \Theta(k)~.
\end{align}
Here $\mc{K}=\{1,2,3,\dots,K\}$ and $K$ is the maximum possible batch size imposed by the underlying queueing system. 
Note that finding the solution of \eqref{optFormulation} runs in $\omega\left(n^4\right)$ time as it 
involves solving $\pmb{\pi} \cdot \mathbf{Q} = 0$~ for every $1 \leq k\leq K$ in \eqref{eq:steadyThroughput}; for a particular batch size $k$,  the dimension of $\mathbf{Q}$ is of order $\frac{n^2}{k}$.
\section{mean-field Model}\label{sec:meanField}

In practical data-processing systems, the number of clients served is usually large.
From a computational point of view, the standard Markovian approach followed in Sect.~\ref{sec:modelSingleJobType} becomes increasingly computationally infeasible when growing the number of clients. 

Consequently, we adopt a mean-field approach where the number of servers~$m$ scales with the number of clients~$n$.
We assume that the batching step is instantaneous, i.e., the number of jobs in the batching station jumps accordingly from $(k-1)$ to $0$ upon the arrival of a new job. This assumption not only simplifies our analysis but is also motivated by empirical observations; for instance, in the commercial database system where we run the evaluation experiments, the batching step is approximately $50$ times faster than the service step.

Let $X^{(n)}(t)$ denote the number of active clients in the system at time $t \geq 0$.
Hence, the number of queries in the system at time $t$ is $n-X^{(n)}(t)$. Then,
$(X^{(n)}(t), t\geq 0)$ is a Markov process with the state space $\cbrac{0,1,\ldots,n}$
and the following rates:
\begin{align*}
q^{(n)}(x \to x-1)&=\lambda x\\
q^{(n)}(x \to x+k)&=\mu(k) \min \brac{m,\floor{\frac{n-x}{k}}}~, 
\end{align*}
where $x\in \cbrac{0,1,\ldots,n}$ and $q(i \to j)$ denotes the transition rate from state $i$ to state $j$.
The Markov process $(X^{(n)}(t), t\geq 0)$ is ergodic because it is 
irreducible and has a finite state space. However, it is extremely difficult to obtain a closed form solution of the
stationary distribution $\pi^{(n)}$ by solving the matrix equation $\pi^{(n)} Q^{(n)}=0$ because of the non-linear
state dependent rates, as mentioned in the previous section. 

An alternative and immediate approach is to obtain a bound on the system throughput as follows. Under the stationary distribution the following must hold:
\begin{equation}
\lambda \expect{X}=k \mu(k) \expect{\min \brac{m,\floor{\frac{n-X}{k}}}}~.
\label{eq:rate_balance}
\end{equation}
Using Jensen's inequality we obtain
\begin{align*}
\lambda \expect{X} \leq k \mu(k) \min \brac{m,\frac{n-\expect{X}}{k}}~,
\end{align*}
which yields the following bound on $\expect{X}$
\begin{equation}
\expect{X} \leq \min \brac{\frac{n \mu(k)}{\lambda+\mu(k)},\frac{k \mu(k)m}{\lambda}}~.
\end{equation}
The throughput of the system is given by the RHS of \eqref{eq:rate_balance}. Hence, an upper bound on the 
throughput $\Theta^{(n)}$ is given by 
\begin{equation}
\expect{\Theta^{(n)}} \leq \min \brac{k \mu(k)m,\frac{n \lambda \mu(k)}{\lambda+\mu(k)}}~.\label{eq:upperboundthroughput}
\end{equation}
(note that we dropped the dependency on $k$ in $\Theta^{(n)}$ for brevity.)

In addition to having this bound on the throughput for finite values of $n$ and $m$, we will next show that the bound is asymptotically tight as $n,m \to \infty$ with $m=\alpha n$ for some fixed $\alpha>0$.

We first consider the process $(w^{(n)}(t), t\geq 0)$, where $$w^{(n)}(t):=X^{(n)}(t)/n$$
denotes the fraction of active clients in the system. The process $(w^{(n)}(t), t\geq 0)$
is a {\em density dependent jump Markov process} \cite{Kurtz_ODE,Mitzenmacher_thesis,ArpanSSY}
with rates
\begin{align*}
q^{(n)}(w \to w-1/n)&=n\lambda w\\
q^{(n)}(w \to w+k/n)&=n\mu(k) \min \brac{\alpha,\frac{1}{n}\floor{\frac{n-nw}{k}}}~,
\end{align*}
where $w:=x/n$. 

Next we prove the following main result:

\begin{theorem}
\label{thm:meanfield_onetype}
\begin{enumerate}[(i)]
\item If $w^{(n)}(0) \to w_0 \in [0,1]$ as $n \to \infty$ in probability,
then we have $$\sup_{0\leq t \leq T} \norm{w^{(n)}(t)-w(t)} \to 0$$
in probability as $n \to \infty$, where $(w(t), t\geq 0)$ is the unique solution of the 
following ODE:
\begin{equation}
\dot w(t) = f(w(t)), \qquad w(0)=w_0,
\end{equation} 
with $f:[0,1] \to \mb{R}$ defined as
\begin{equation}
f(w)= k \mu(k) \min \brac{\alpha, \frac{1-w}{k}} -\lambda w~.
\end{equation}

\item For any $w_0 \in [0,1]$, we have $w(t) \to w^*$ exponentially fast as $t \to \infty$, where
$w^*$ is the unique solution of $f(w^*)=0$ and is given by
\begin{equation}
w^*=\min \brac{\frac{\mu(k)}{\lambda+\mu(k)}, \frac{\alpha k \mu(k)}{\lambda}}
\label{eq:fixed_point}
\end{equation}
\item The sequence of stationary measures $\pi_w^{(n)}$ of the process\\ $(w^{(n)}(t), t \geq 0)$
converges weakly to $\delta_{w^*}$ as $n \to \infty$. 
\end{enumerate}
\end{theorem}

\begin{proof}
To show part (i), we first note that the limiting expected drift of the
process $(w^{(n)}(t),t\geq 0)$ conditioned on $w^{(n)}(t)=w$ converges point-wise (and hence uniformly)
to the continuous function $f$,
i.e., for each $w \in [0,1]$ we have

\begin{equation}
\lim_{n \to \infty} \lim_{h \to 0} \frac{1}{h}\expect{w^{(n)}(t+h)-w^{(n)}(t) \vert w^{(n)}(t)=w}=f(w)~.
\end{equation} 
Furthermore, it is easy to see that $f:[0,1] \to \mb{R}$ is Lipschitz continuous which follows from
the facts (1) any linear function is Lipschitz continuous, (2) if $F,G$ are Lipschitz continuous, then
$cF+dG$ is Lipschitz continuous for any $c,d \in \mb{R}$, (3) $\abs{F}$ is Lipschitz continuous when
$F$ is Lipschitz continuous, and (4) $\min(F,G)=\frac{F+G}{2}-\frac{\abs{F-G}}{2}$.
Part (i) now follows from Theorem 3.1 of \cite{Kurtz_ODE}.

To prove part (ii), we first observe that the unique solution to the equation $f(w^*)=0$
is given by \eqref{eq:fixed_point}. Without loss of generality we assume that $w_0 \geq w^*$.
Then $w(t) \geq w^*$ for all $t \geq 0$ due to the continuity of the process $w(t)$ and the fact that $\dot w(t)=0$
when $w(t)=w^*$. We define the distance function $\phi(t)=w(t)-w^*$. Clearly, $\phi(t) \geq 0$ for all $t \geq 0$.
Now we have
\begin{align*}
\dot{\phi}(t)=\dot{w}(t)&=f(w)\\
&=f(w)-f(w^*)\\
&=-\lambda(w-w^*)+\\
& \quad k \mu(k) \sbrac{\min\brac{\alpha, \frac{1-w(t)}{k}}-\min\brac{\alpha, \frac{1-w^*}{k}}} \\
&\leq -\lambda \phi,
\end{align*}
where the last inequality follows since $w(t) \geq w^*$ for all $t \geq 0$.
From the above we see that $\phi(t) \leq \phi(0) e^{-\lambda t}$. This implies that
$$w(t) \to w^*$$ as required.

To show part (iii), we first note that the stationary measure $\pi_w^{(n)}$ is tight
as it is defined on the compact space $[0,1]$. Hence, part (iii) follows from 
Theorem 2 of \cite{Gast_chap}.
\end{proof}

The above theorem implies the weaker result that
$$\lim_{n \to \infty} \lim_{t \to \infty}\expect{w^{(n)}(t)}=\lim_{t\to \infty} \lim_{n\to \infty} \expect{w^{(n)}(t)}=w^*.$$
Equivalently, we have the following convergence of the normalized throughput $\Theta^{(n)}/n$
$$\lim_{n \to \infty} \expect{\Theta^{(n)}/n}=\lambda w^*,$$
which proves the asymptotic tightness of the bound from (\ref{eq:upperboundthroughput}).

The optimal asymptotic throughput further follows by maximizing the fraction of active clients $w^*$ with respect to 
the batch size $k$. The asymptotically optimal batch size is the solution to the following optimization problem
\begin{equation}
\underset{k}{\max}  \min \brac{\frac{\mu(k)}{\lambda+\mu(k)}, \frac{\alpha k \mu(k)}{\lambda}} \label{eq:k_opt}~.
\end{equation}
In the particular case when $\mu(k)$ is a non-increasing function of $k$ and $k \mu(k)$ is a non-decreasing function
of $k$, the optimal solution $k^*$ is simply the solution to the following equation
\begin{equation}
\frac{\mu(k)}{\lambda+\mu(k)}= \frac{\alpha k \mu(k)}{\lambda}~.
\label{eq:optimal_k_mean_field}
\end{equation}

Therefore, we have just showed that $k^*$ can simply be found by solving a polynomial equation.
We can approximate the optimal batch size for finite systems by $k^*$
as long as $n$ and $m$ are large. The advantage is that solving the polynomial equation can be done in time independent of the system size $n$; moreover, as we will show in our numerical experiments, the approximation is numerically accurate in practical regimes.
\section{The Two Job-Type Case}\label{sec:modelMultJobTypeBig}

\subsection{Queueing Model and Exact Solution}\label{sec:modelMultJobType}
We now consider the case when jobs can be of two types, e.g., \textit{write} and \textit{read} in a database system. Each of these types benefits from batching and can possibly have different speedups; we note that batching involves jobs of the same type, which is typically the case in database systems.
Additionally, we consider priority service scheduling between the two types, which can be either preemptive or non-preemptive. In a database system, where queries can be of type \textit{write} or \textit{read}, the former is usually prioritized.

In our model, we assume without loss of generality that the first type is given priority in the service station. Below we describe the system dynamics and the required state space representation before providing the mean-field formulation. 

Recall that the producer station has $n$ clients, each producing one job with rate $\lambda$ once becoming active (i.e., once receiving their token back); also, the number of active clients is denoted by $x$.
In the two job-type model, each \emph{active} client produces a job of type $1$ with probability $p$ or a job of type $2$ with probability $(1-p)$. 
The number of type $1$ and type $2$ jobs in the batching station is denoted by $y_1$ and $y_2$, respectively. The batching station groups $k_i$ jobs of type $i$ into a batch  with rate $M_i(k_i)\floor{y_i/k_i}$ whenever $y_i \ge k_i$, $i \in \{1,2\}$, and forwards batches to the service station. Further, the service station has $m$ parallel servers which give \textit{preemptive priority} to the type $1$ jobs; the alternative case of non-preemptive priority is discussed in Sect. ~\ref{sec:app-A} of the Appendix.

Let us denote the total number of type $1$ batches by $z_1$. Due to preemptive priority, the actual number of type $1$ batches in service is $v_1 = \min(m,z_1)$. The rest of the servers may be occupied by batches of type $2$.
The state of the system can be uniquely described by the quadruple $(x,y_1,y_2,z_1 k_1)$, where $(x,y_1,y_2,z_1 k_1)$ belongs to the state space 
\begin{align*}
    \mc{S}=\cbrac{(x_1,x_2,x_3,x_4):\in \mb{Z}_+^4: x_1+x_2+x_3+x_4 \le n, k|x_4}. 
\end{align*}
Note that the number of type $2$ jobs in the system which are already batched is $$z_2 k_2 = (n-x-y_1-y_2-z_1 k_1)~,$$ out of which $v_2 k_2$ are at the server and the rest are queued for service; here,   
\begin{align}
\label{eq:type2batchesInServicePreemptive}
    v_2 = \min(\max(0,m-z_1),z_2)~. 
\end{align}
Clearly, the system evolves as a continuous-time Markov chain with the jump rates 
\begin{align}
\label{eq:jumpRatesPreemptive}
    s &\xrightarrow[\text{}]{\lambda x p} s-\mathbf{e_1}+\mathbf{e_2}, \thinspace x>0 \nonumber \\
    &\xrightarrow[\text{}]{\lambda x (1-p)} s-\mathbf{e_1}+\mathbf{e_3}, \thinspace x>0 \nonumber \\
            &\xrightarrow[\text{}]{M_1(k_1)\floor{y_1/k_1}} s-k_1\mathbf{e_2}+k_1\mathbf{e_4}, \thinspace y_1 \ge k_1 \nonumber \\
            &\xrightarrow[\text{}]{v_1\mu_1(k_1)} s+k_1\mathbf{e_1}-k_1\mathbf{e_4}, \thinspace z_1 \ge 1 \nonumber\\
            &\xrightarrow[\text{}]{v_2\mu_2(k_2)} s+k_2\mathbf{e_1}, \thinspace v_2 \ge 1~,
\end{align}
where $s=(x,y_1,y_2,z_1 k_1)$ and $e_j$ is the unit vector of appropriate size whose $j$-th component is unity. The jump rates to all the other state are zero.

The chain is irreducible whereas the finiteness of the state space guarantees positive recurrence. Thus, we can derive the rate matrix $\mathbf{Q}$ using ~\eqref{eq:jumpRatesPreemptive} and derive the steady state distribution $\pmb{\pi}_0$ by solving $\pmb{\pi} \cdot \mathbf{Q} = 0$. 

While we could jointly optimize for $k_1$ and $k_2$, database batching argues for using a uniform batch size across all job types (see, e.g.,~\cite{Sellis:1988:MO:42201.42203:MQO,giannikis2012shareddb,rehrmann2018oltpshare}); in particular, standard multi-query optimization methods in databases batch requests through fixed compiling of the execution of multiple queries into one SQL string which renders equal batch sizes regardless of type. Denoting $k:=k_1=k_2$, the steady state throughput is
\begin{align}\label{eq:steadyThroughputPreemptive}
    \Theta_p(k) = \sum_{s \in \mc{S}} \pmb{\pi}_0(s) k(\mu_1(k) v_1+\mu_2(k) v_2)~,
\end{align}
where $s = (x,y_1,y_2,z_1k)$ and $v_2$ is derived in ~\eqref{eq:type2batchesInServicePreemptive}.
The optimal batch size is 
\begin{align}
\label{optFormulationPreemptive}
    k^* = \argmax_{k \in \mc{K}} \Theta_p(k)~.
\end{align}
Here, $\mc{K}=\{1,2,3,\dots,K\}$ and $K$ is the maximum possible batch size for the considered system.

\subsection{Mean-field Formulation: Preemptive Priority}\label{subsec:mFTwoJobs}

We now discuss the preemptive priority case in the context of the mean-field formulation from Sect.~\ref{sec:meanField}. 
The system can now be uniquely described by the number of active clients and the number of type $1$ jobs in the system. This is due to the fact that there can be at most $(k_1-1)$ jobs of type $1$ that have not formed a batch; the number of un-batched jobs is $mod(x_2,k_1)$, where $x_2$ is number of type~$1$ jobs in the system. This phenomenon also applies to the type~$2$ jobs and lets us derive the number of type~$2$ jobs which are not yet batched. Assuming work conservingness of the server and the preemptive priority of type $1$ over type $2$, we can derive the number of batches in service for each type. Note that we use the notation $\mu_1$ and $\mu_2$ instead of $\mu_1(k_1)$ and $\mu_2(k_2)$ when the dependence is clear. 

Let $X_1^{(n)}(t)$ and $X_2^{(n)}(t)$ denote the numbers of active clients and the total number of 
type 1 jobs in the system at time $t \geq 0$, respectively.
Then at time $t \geq 0$, the number of type 2 jobs in the system is $n-X_1^{(n)}(t)-X_2^{(n)}(t)$,
the number of type 1 batches being served is $\min\brac{m,\floor{X_2^{(n)}(t)/k_1}}$, and the 
number of type 2 batches being served is $$\min\brac{m-\min\brac{m,\floor{X_2^{(n)}(t)/k_1}}, \floor{(n-X_1^{(n)}(t)-X_2^{(n)}(t))/k_2}},$$ which simplifies to $$\min\brac{\max\brac{0,m-\floor{X_2^{(n)}(t)/k_1}}, \floor{(n-X_1^{(n)}(t)-X_2^{(n)}(t))/k_2}}.$$
Clearly, $(X_1^{(n)}(t),X_2^{(n)}(t), t\geq 0)$ is Markov process on state space $\mc{S}=\cbrac{(x_1,x_2)\in \mb{Z}_+^2: x_1+x_2 \leq n}$ with the following rates:

\begin{align*}
&q((x_1,x_2) \to (x_1-1,x_2+1))= \lambda p x_1\\
&q((x_1,x_2) \to (x_1-1,x_2))= \lambda (1-p) x_1\\
&q((x_1,x_2) \to (x_1+k_1,x_2-k_1))= \mu_1 \min\brac{m,\floor{\frac{x_2}{k_1}}}\\
&q((x_1,x_2) \to (x_1+k_1,x_2-k_1))= \mu_1 \min\brac{m,\floor{\frac{x_2}{k_1}}}\\
&q((x_1,x_2) \to (x_1+k_2,x_2))= \\ &\qquad \mu_2 \min\brac{\max\brac{0,m-\floor{\frac{x_2}{k_1}}},\floor{\frac{n-x_1-x_2}{k_2}}}\\
\end{align*} 
As in the previous section, we consider the scaled process $w^{(n)}(t)=(w_1^{(n)}(t), w_2^{(n)}(t))$ 
with $w_i^{(n)}(t)=X_i^{(n)}(t)/n$, $i=1:2$. We show the following theorem

\begin{theorem}\label{thm:meanFieldTwoJobs}
\begin{enumerate}[(i)]
\item If $w^{(n)}(0) \to w_0 \in [0,1]^2$ as $n \to \infty$ in probability,
then we have $$\sup_{0\leq t \leq T} \norm{w^{(n)}(t)-w(t)} \to 0$$
in probability as $n \to \infty$, where $(w(t)=(w_1(t),w_2(t)), t\geq 0)$ is the unique solution of the 
following system of ODEs:

\begin{equation}
\dot w_1(t) = f_1(w(t)), \qquad \dot w_2(t) = f_2(w(t)), \quad w(0)=w_0,
\end{equation} 
with $f=(f_1,f_2):[0,1]^2 \to \mb{R}^2$ defined as
\begin{align}
f_1(w) &= - \lambda w_1+k_1 \mu_1 \min \brac{\alpha, \frac{w_2}{k_1}}+ \nonumber \\
& \qquad k_2\mu_2 \min\brac{\max\brac{0,\alpha-\frac{w_2}{k_1}},\frac{1-w_1-w_2}{k_2}}  \\
f_2(w) &= \lambda p w_1- k_1 \mu_1 \min \brac{\alpha, \frac{w_2}{k_1}} 
\end{align}

\item For any $w_0 \in [0,1]^2$, we have $w(t) \to w^*$ as $t \to \infty$, where
$w^*=(w_1^*,w_2^*)$ is the unique solution of $f(w^*)=0$ and is given by

\begin{align}
w_1^* & = \min \brac{\frac{\mu_1 \mu_2}{\mu_1  \lambda (1-p)+\mu_2 \lambda p+ \mu_1 \mu_2}, \frac{k_1 k_2 \mu_1 \mu_2 \alpha}{k_1 \mu_1  \lambda (1-p)+k_2 \mu_2 \lambda p}}\\
w_2^* &=\frac{\lambda p w_1^*}{\mu_1}
\label{eq:fixed_point_two_types}
\end{align}

\item The sequence of stationary measures $\pi_w^{(n)}$ of the process\\ $(w^{(n)}(t), t \geq 0)$
converges weakly to $\delta_{w^*}$ as $n \to \infty$. 
\end{enumerate}
\end{theorem}
\begin{proof}
Part (i) can be shown using arguments similar to the proof of Part (i)
of Theorem~\ref{thm:meanfield_onetype}.
To show part (ii), we first note that $w^*$ is the unique solution of $f(w^*)=0$.
We now show that $w^*$ is globally attractive. 

We first define a linear transform 
$(w_1,w_2) \to (z_1,z_2)$ defined as $z_1=w_1+w_2$ and $z_2=w_2$. Under this transformation
the system is described as follows:

\begin{align}
    \frac{dz_1}{dt}&=\begin{cases}
                        -\lambda (1-p)(z_1-z_2), \quad \text{if } z_2 \geq k_1 \alpha\\
                        -\lambda (1-p)(z_1-z_2)+\mu_2(1-z_1), \quad \text{if } \frac{1-z_1}{k_2}+\frac{z_2}{k_1}<\alpha\\
                        -\lambda (1-p)(z_1-z_2)-\frac{k_2}{k_1}\mu_2 z_2 +k_2 \mu_2 \alpha, ~ \text{if } \frac{1-z_1}{k_2}+\frac{z_2}{k_1} \ge \alpha
                    \end{cases}\\
    \frac{dz_2}{dt}&=\begin{cases}
                        \lambda p (z_1-z_2)-k_1 \mu_1 \alpha, \quad \text{if } z_2 \geq k_1 \alpha\\
                        \lambda p(z_1-z_2) -\mu_1 z_2, \quad \text{otherwise}.
                    \end{cases}
\end{align}
Furthermore, the stationary point is mapped to $(z_1^*,z_2^*)$, where $z_1^* = \min (z_{11}^*,z_{12}^*)$ and $z_2^* =\eta z_1^*$ with 
$z_{11}^*  =  \frac{(\mu_1+ \lambda p) \mu_2}{\mu_1  \lambda (1-p)+\mu_2 \lambda p+ \mu_1 \mu_2}$, 
$z_{12}^*  = \frac{k_1 k_2 (\mu_1+ \lambda p) \mu_2 \alpha}{k_1 \mu_1  \lambda (1-p)+k_2 \mu_2 \lambda p}$, 
$\eta  = \frac{\lambda p }{\mu_1+ \lambda p}$.

Clearly, the system is a piece-wise linear system. We consider the 
stability of each region individually:

{\bf Case 1}: $k_1 \alpha \geq 1$, $k_2 \alpha \geq 1$\\
In this case, the system reduces to the following system 
(since the domain of interest is $1\geq z_1 \geq z_2 \geq 0$):
\begin{align}\label{eq:evolutionTwoTypesCase1}
    \frac{dz_1}{dt}&=-\lambda(1-p)(z_1-z_2)+\mu_2(1-z_1) \nonumber\\
    \frac{dz_2}{dt}&=\lambda p(z_1-z_2)-\mu_1z_2
\end{align}

The above system can be represented as a linear dynamical system $\dot z=Az+b$,
where the eigenvalues $\theta$ of $A \in \mb {R}^{2\times 2}$ satisfy 
$$\theta^2+(\lambda + \mu_1 + \mu_2)\theta+c_0=0,$$
for some constant $c_0$. Clearly, the real parts of the eigenvalues are strictly negative. Hence, the system
is globally attractive to the unique stationary point 
$(z_{11}^*, \eta z_{11}^*)$.

{\bf Case 2}: $k_1 \alpha \geq 1$, $k_2 \alpha < 1$\\
In this case, we see that the domain of interest is divided into two regions by the line 
\begin{align*}
 L_1: z_2 &=\frac{k_1}{k_2}(k_2 \alpha -1 + z_1),
\end{align*}
having respective linear equations. Let us also consider the line
\begin{align*}
 L_2: z_2 & = \eta z_1.
\end{align*}
Note that $\dot z_2(t) \ge 0$ iff $z$ lies below $L_1$. Thus, if the system starts below $L_1$, it stays there and vice-versa and the fixed point(s) of the system, if exist(s), lie(s) on $L_2$. Let $z_{10}$ denote $z_1$-coordinate of the intersection point of these lines, i.e., 
\begin{align*}
z_{10} = \frac{k_1 (\mu_1+ \lambda p) (1- k_2 \alpha)}{k_1 \mu_1 +k_1 \lambda p - k_2 \lambda p}.
\end{align*}
Let us assume 
\begin{align*}
z_{11}^* \le z_{12}^*.    
\end{align*}
Calculations show that this implies $z_{10} \le z_{11}^* \le z_{12}^*$. If the system starts from a point below $L_1$, the evolution of the system is given by \eqref{eq:evolutionTwoTypesCase1} and a similar argument as \emph{Case 1} shows that the system converges to the fixed point $(z_{11}^*, \eta z_{11}^*)$. In case the initial point lies above $L_1$, the evolution in the starting phase is given by
\begin{align}
\label{eq:evolutionTwoTypesCase2New}
    \frac{dz_1}{dt}&=-\lambda(1-p)(z_1-z_2)+ k_2 \mu_2(\alpha -\frac{z_2}{k_1}) \nonumber\\
    \frac{dz_2}{dt}&=\lambda p(z_1-z_2)-\mu_1z_2
\end{align}
Calculations show that the real parts of the corresponding eigenvalues are negative and the fixed point for this system is given by $(z_{12}^*, \eta z_{12}^*)$. Thus the system crosses $L_1$ where the evolution is governed by \eqref{eq:evolutionTwoTypesCase1}. From the perspective of convergence, this is equivalent to having the initial point below $L_1$ in which is case the convergence to $(z_{11}^*, \eta z_{11}^*)$ is already established. Thus, the system always converges to $(z_{11}^*, \eta z_{11}^*)$ when $z_{11}^* \le z_{12}^*$. 

For the case
\begin{align*}
z_{11}^* > z_{12}^*,
\end{align*}
we notice that $z_{10} \ge z_{11}^* \ge z_{12}^*$ and a similar argument shows convergence of the system to $(z_{12}^*, \eta z_{12}^*)$.

{\bf Case 3}: $k_1 \alpha < 1$, $k_2 \alpha < 1$\\
In this case, we can have different possibilities for the initial state.
We first consider the case where we start with a vector $(z_1,z_2)$
such that $1\geq z_1 \geq z_2 \geq k_1 \alpha$. We will show that the system eventually 
reaches a state where $z_2 \leq k_1 \alpha$. In this case, until we have $z_1 \leq k_1 \alpha$,
the evolution is given by
\begin{align}\label{eq:evolutionTwoTypesCase2Old}
    \frac{dz_1}{dt}&=-\lambda (1-p)(z_1-z_2) \nonumber\\
    \frac{dz_2}{dt}&=\lambda p (z_1-z_2)-k_1 \mu_1 \alpha
\end{align}
The above is clearly a unstable system with $z_1$ and $z_2$ decreasing indefinitely for ever.
Therefore, there exists $t \geq 0$ such that $z_2(t) \leq k_1 \alpha$.

Now without loss of generality we start our system with $z_2 \leq k \alpha$. Thus, without loss of generality, we take an initial point satisfying $z_2 \le k_1 \alpha$. Let us assume
\begin{align*}
\eta \le k_1 \alpha,
\end{align*}
Like \emph{Case 2}, we observe that either $z_{10} \le z_{11}^* \le z_{12}^*$ or $z_{10} \ge z_{11}^* \ge z_{12}^*$ and the proof follows the same line of argument as \emph{Case 2}.

Now let's consider the scenario when
\begin{align*}
\eta > k_1 \alpha.
\end{align*}
We show that $z_{11}^* \ge z_{12}^*$ which holds if and only if
\begin{align*}
\mu_1 \mu_2 k_1 k_2 \alpha \le k_1 \mu_1  \lambda (1-p) (1- k_2 \alpha) +k_2 \mu_2 \lambda p (1- k_1 \alpha).
\end{align*}
Since, $\eta > k_1 \alpha$, it suffices to show
\begin{align*}
\mu_1 \mu_2 k_2 \eta \le k_1 \mu_1  \lambda (1-p) (1- k_2 \alpha) +k_2 \mu_2 \lambda p (1- \eta),
\end{align*}
which is equivalent to $k_1 (\mu_1 + \lambda p) (1-p) (1- k_2 \alpha) \ge 0$. Similar to \emph{Case 2}, we see that if the initial point lies above $L_1$, the evolution is given by \eqref{eq:evolutionTwoTypesCase2New} and the system converges to $(z_{12}^*, \eta z_{12}^*)$. When started  below $L_1$, the evolution is governed by \eqref{eq:evolutionTwoTypesCase1} initially and the system moves towards $(z_{11}^*, \eta z_{11}^*)$. This eventually changes the evolution dynamics to \eqref{eq:evolutionTwoTypesCase2New} and the system converges to $(z_{12}^*, \eta z_{12}^*)$ in either case.

{\bf Case 4}: $k_1 \alpha < 1$, $k_2 \alpha \ge 1$\\
Using the same argument as \emph{Case 2}, we take an initial point with $z_2 \le k_1 \alpha$. 

Let us first assume
\begin{align*}
\eta \le k_1 \alpha.
\end{align*}
Similar to the argument of \emph{Case 3} when $\eta > k_1 \alpha$ and using the fact that $k_2 \alpha \ge 1$, we observe this implies $z_{11}^* \le z_{12}^*$. The convergence from an initial point below or above $L_1$ follows in a similar fashion.

The remaining scenario is  
\begin{align*}
\eta > k_1 \alpha.
\end{align*}
Similar to \emph{Case 2}, we observe that either $z_{10} \le z_{11}^* \le z_{12}^*$ or $z_{10} \ge z_{11}^* \ge z_{12}^*$ and convergence can be shown to $(z_{11}^*, \eta z_{11}^*)$ or $(z_{12}^*, \eta z_{12}^*)$, respectively, using the same line of argument presented there.
Thus global attraction is established under all scenarios. Note that we have mentioned the actual limit as $(z_{11}^*, \eta z_{11}^*)$ or $(z_{12}^*, \eta z_{12}^*)$, as applicable.

Finally Part (iii) of the theorem follows by the same line arguments as in the proof of Part (iii)
of Theorem~\ref{thm:meanfield_onetype}. A more general result under the assumption of equal batch sizes is given in Appendix \ref{sec:manyJobManyService}.
\end{proof}

\begin{figure*}[h!]
	\centering
	\begin{subfigure}[t]{0.32\textwidth}
		\centering
		\includegraphics[width=1\textwidth]{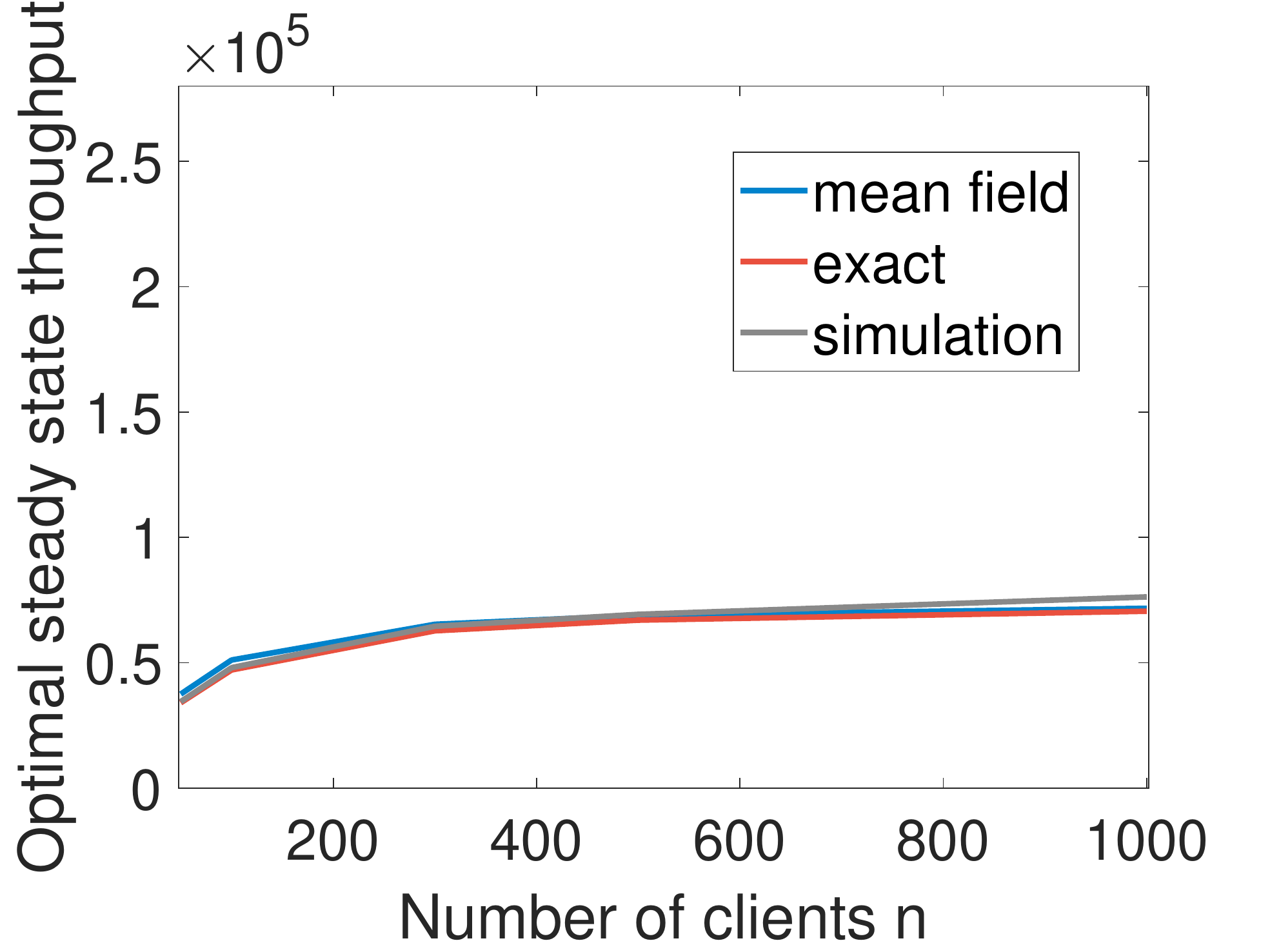}
		\caption{\label{fig:optTCompare4servSingle}%
		$m=4$ (servers)}
	\end{subfigure}
    \hfill
	\begin{subfigure}[t]{0.32\textwidth}
		\centering
		\includegraphics[width=1\textwidth]{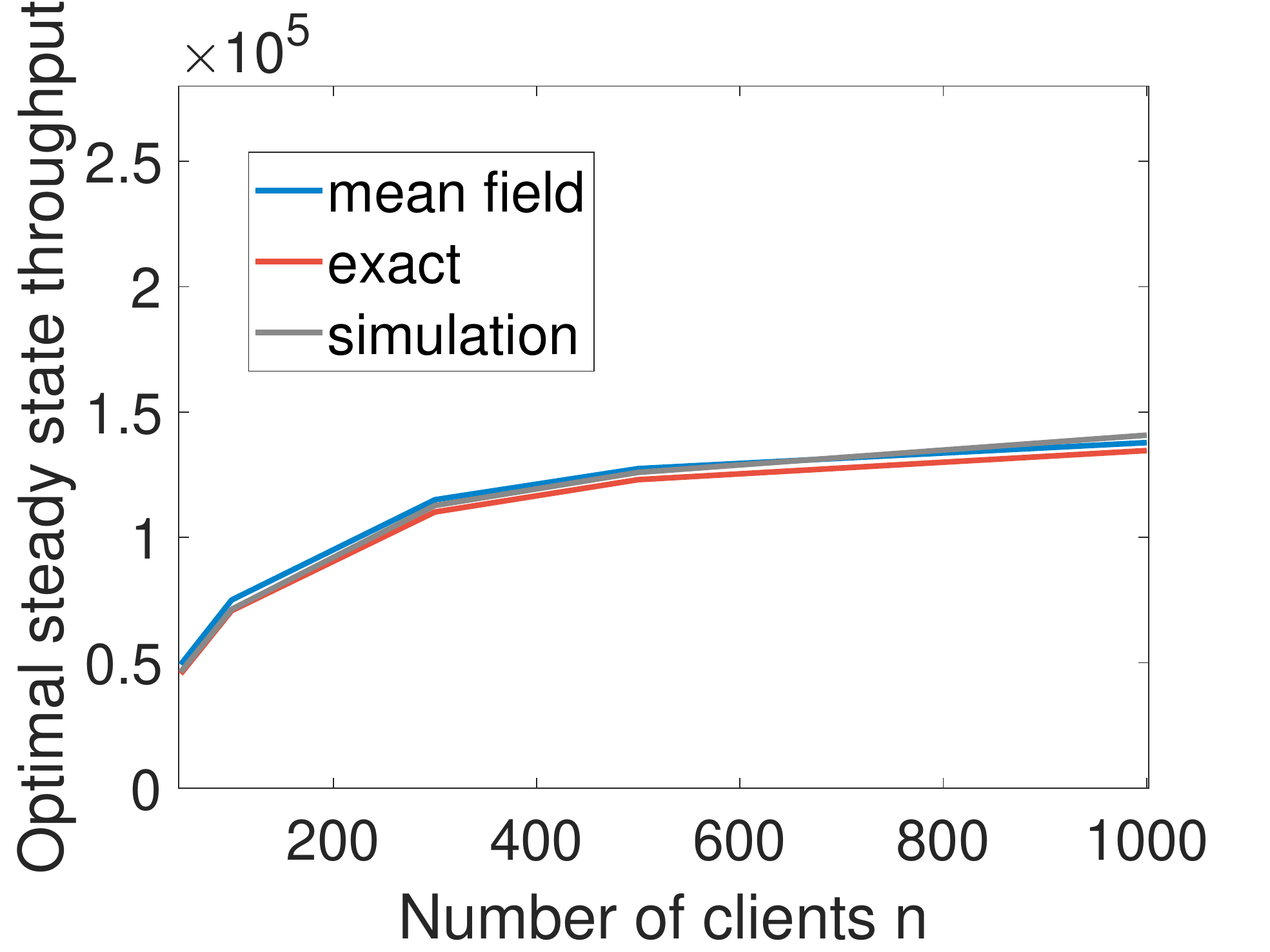}
		\caption{\label{fig:optTPutCompare8servSingle}%
        $m=8$}
	\end{subfigure}
    \hfill
	\begin{subfigure}[t]{0.32\textwidth}
		\centering
		\includegraphics[width=1\textwidth]{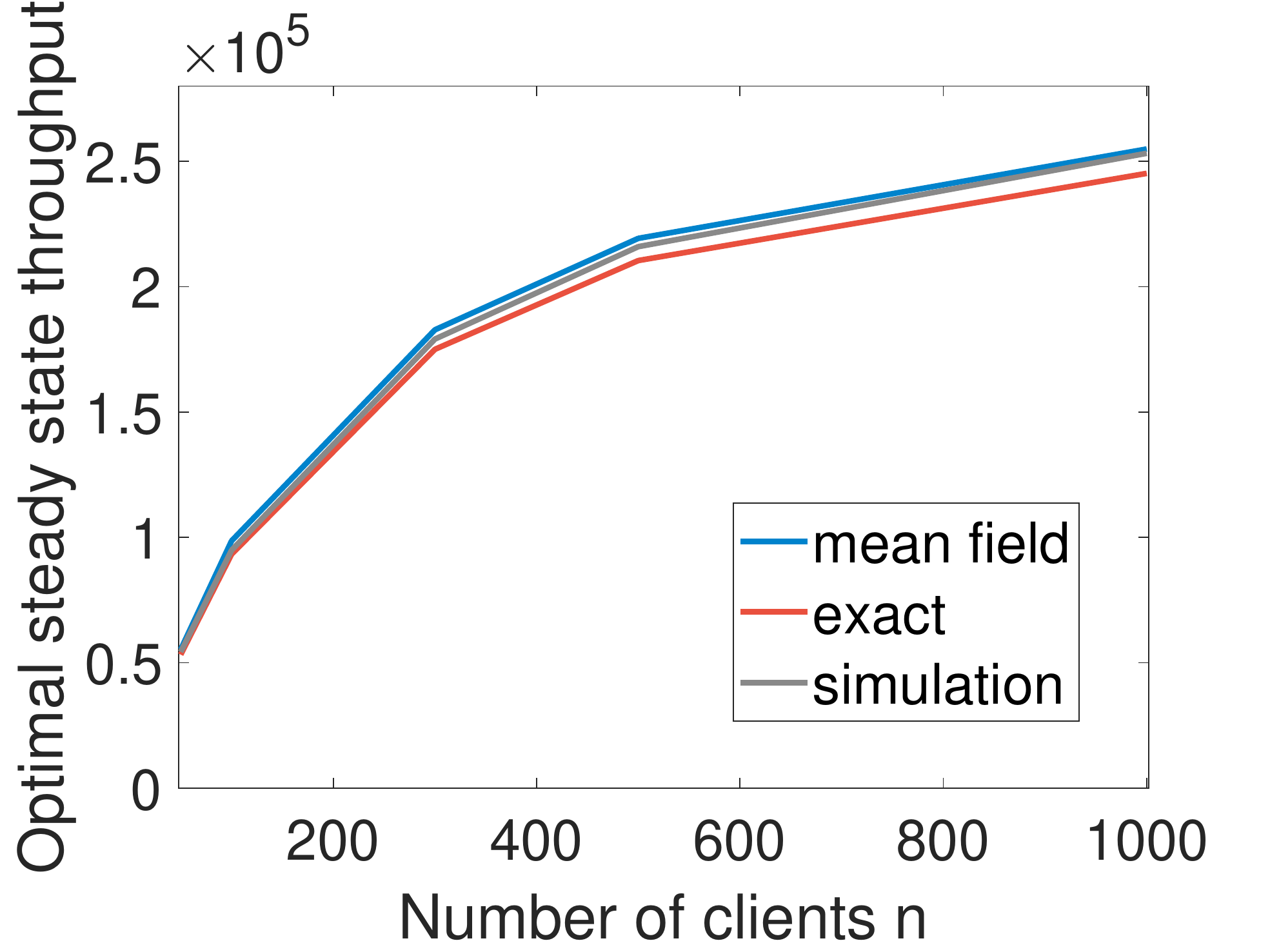}
		\caption{\label{fig:optTCompare16servSingle}%
		$m=16$}
	\end{subfigure}
\caption{\label{fig:optTPut-vs-n-1job}%
The optimal steady state throughput as a function of the number of clients/jobs $n$, for the single job type case; results from the mean-field model, the non-asymptotic/exact formulation, and simulations, for several values of the number of servers $m$ and a linear service speedup. For a fixed $m$, the optimal throughput is known from the mean-field analysis to be $m k^* \mu(k^*)$, where $k^*$ is given in~\eqref{eq:optimal_k_mean_field}.}%
\end{figure*}

\begin{figure*}[h!]
	\centering
	\begin{subfigure}[t]{0.32\textwidth}
		\centering
		\includegraphics[width=1\textwidth]{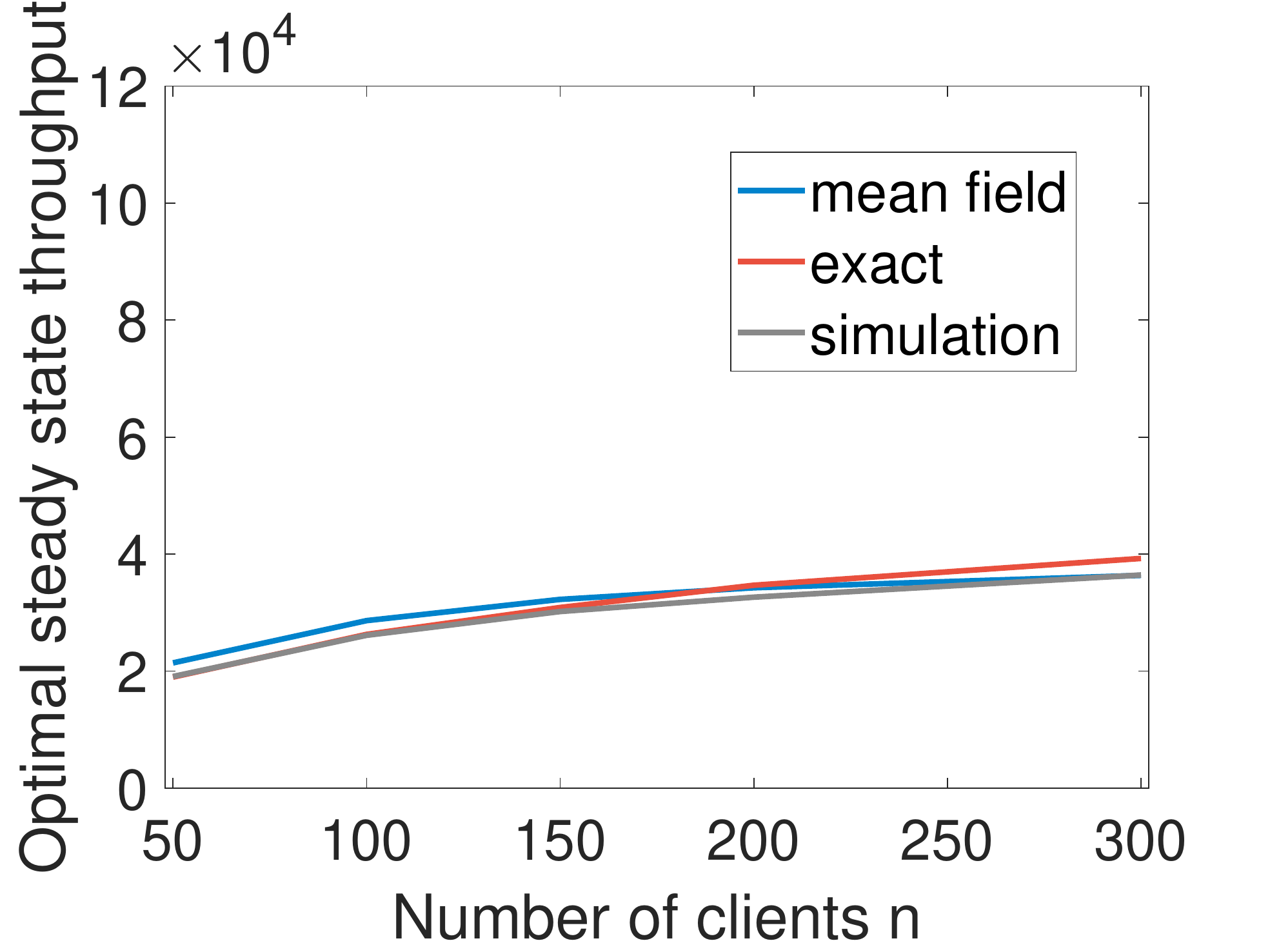}
		\caption{\label{fig:optTCompare4servPreEmp}%
		$m=4$}
	\end{subfigure}
    \hfill
	\begin{subfigure}[t]{0.32\textwidth}
		\centering
		\includegraphics[width=1\textwidth]{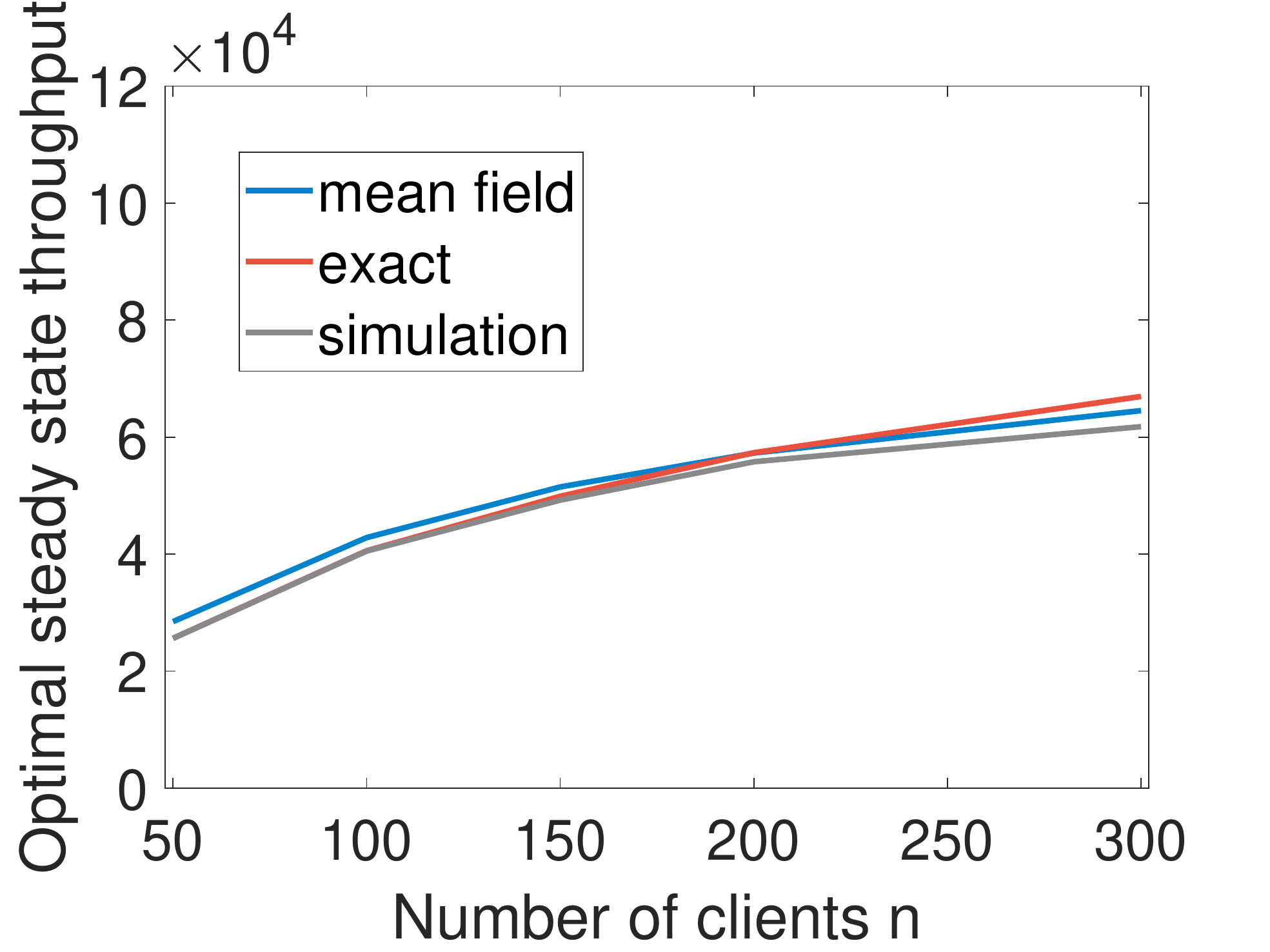}
		\caption{\label{fig:optTPutCompare8servPreEmp}%
        $m=8$}
	\end{subfigure}
    \hfill
	\begin{subfigure}[t]{0.32\textwidth}
		\centering
		\includegraphics[width=1\textwidth]{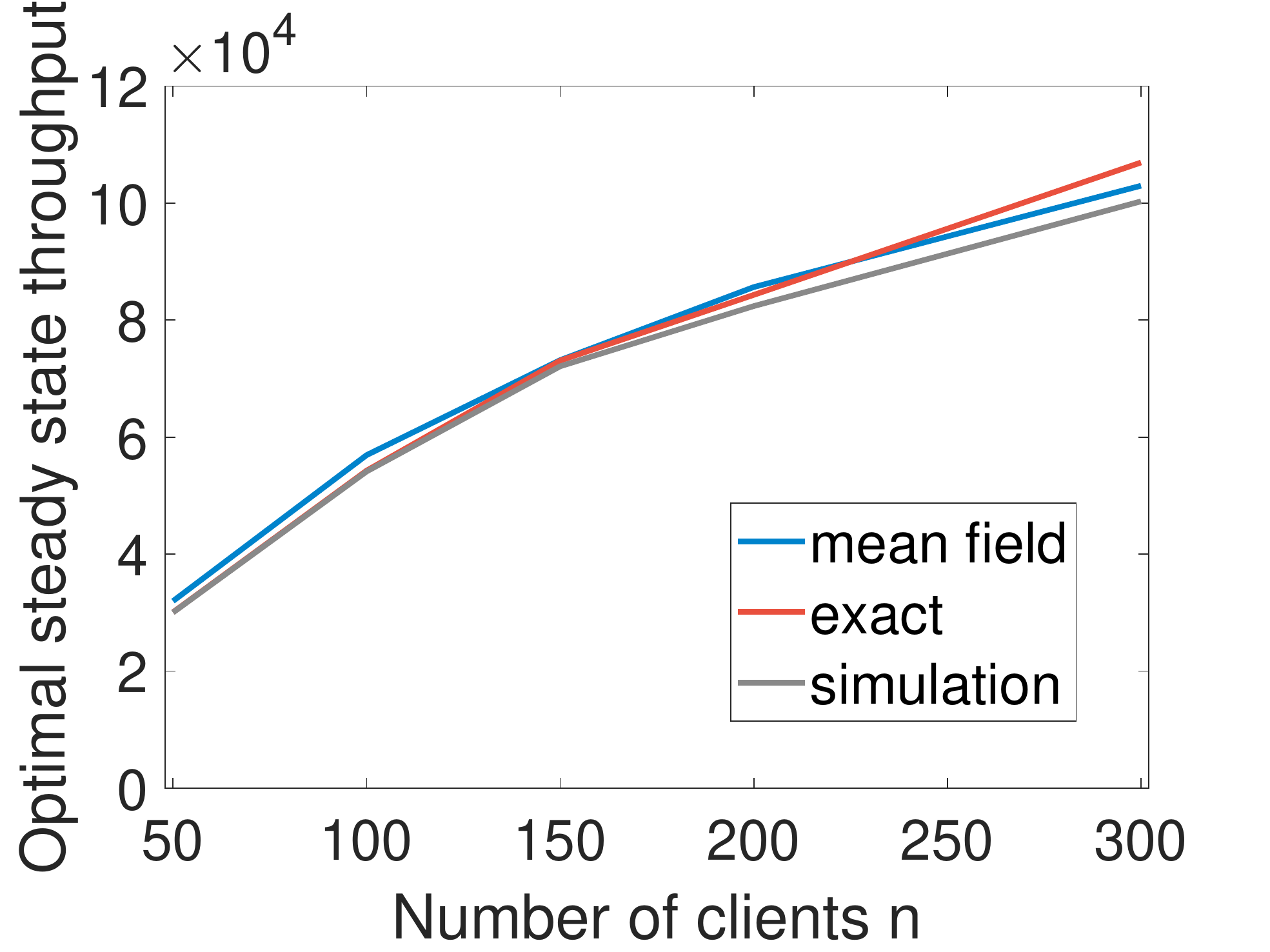}
		\caption{\label{fig:optTCompare16servPreEmp}%
		$m=16$}
	\end{subfigure}
\caption{\label{fig:optTPut-vs-n-2jobs}%
The optimal total steady state throughput for the two-job types case, preemptive priority, and linear speedup.}%
\end{figure*}

From the above theorem it follows that the asymptotic throughput is a linear combination
of $w_1^*$ and $w_2^*$. Given the forms of the speedup functions $\mu_1(k)$ and $\mu_2(k)$,
we can optimize the asymptotic throughput jointly over $k_1$ and $k_2$. 
The time taken to find the asymptotic optimal
batch sizes is clearly independent of the system size $n$ and these asymptotic solutions
serve as accurate estimates for batch sizes for finite systems, as we will show in Sec.~\ref{sec:numerical}.

\section{Evaluation}
\label{sec:numerical}
In this section, we evaluate the performance of our model for throughput optimization using both simulations and an application to a research prototype of a large commercial database system.
We first show accuracy of our model for simulation results and subsequently describe the details of experimental evaluations which includes the system layout, the experiment description, data collection and the model performance.

\subsection{Simulations} \label{sec:numericalEval_}
We first numerically compare our exact and asymptotic results to corresponding simulation results. For all comparisons, the \textit{exact} model obtains the throughput by solving for the steady state distribution numerically whereas for \textit{simulations}, we plot the observed throughput when the system is simulated using \eqref{eq:jumpRates}. The unit of time for simulations is seconds and a  linear form of speedup is assumed. Further, the following system parameters are used for the single job type case (The parameters correspond to the range of values observed in the prototype system described in the next section):
\begin{itemize}
    \item job generation rate $\lambda = 5\cdot 10^{3}$,
    \item batch service time $1/\mu(k) = 3.6\cdot 10^{-4}+5.2\cdot 10^{-5} ~k$,
    \item batching time $1/M(k) = 7.2\cdot 10^{-6}+1\cdot 10^{-6} ~k$.
\end{itemize}
For two job types, type $1$ job has higher priority and is generated with $20\%$ probability. We modify the service rates as below and keep other parameters unchanged.
\begin{itemize}
    \item type $1$ service time $1/\mu_1(k) = 1/(5 \cdot \mu_2(k))$
     \item type $2$ service time $1/\mu_2(k) = 5.4\cdot 10^{-4}+5.3\cdot 10^{-4} ~k$
\end{itemize}

\begin{figure*}[t!]
	\centering
	\begin{subfigure}[t]{0.32\textwidth}
		\centering
		\includegraphics[width=1\textwidth]{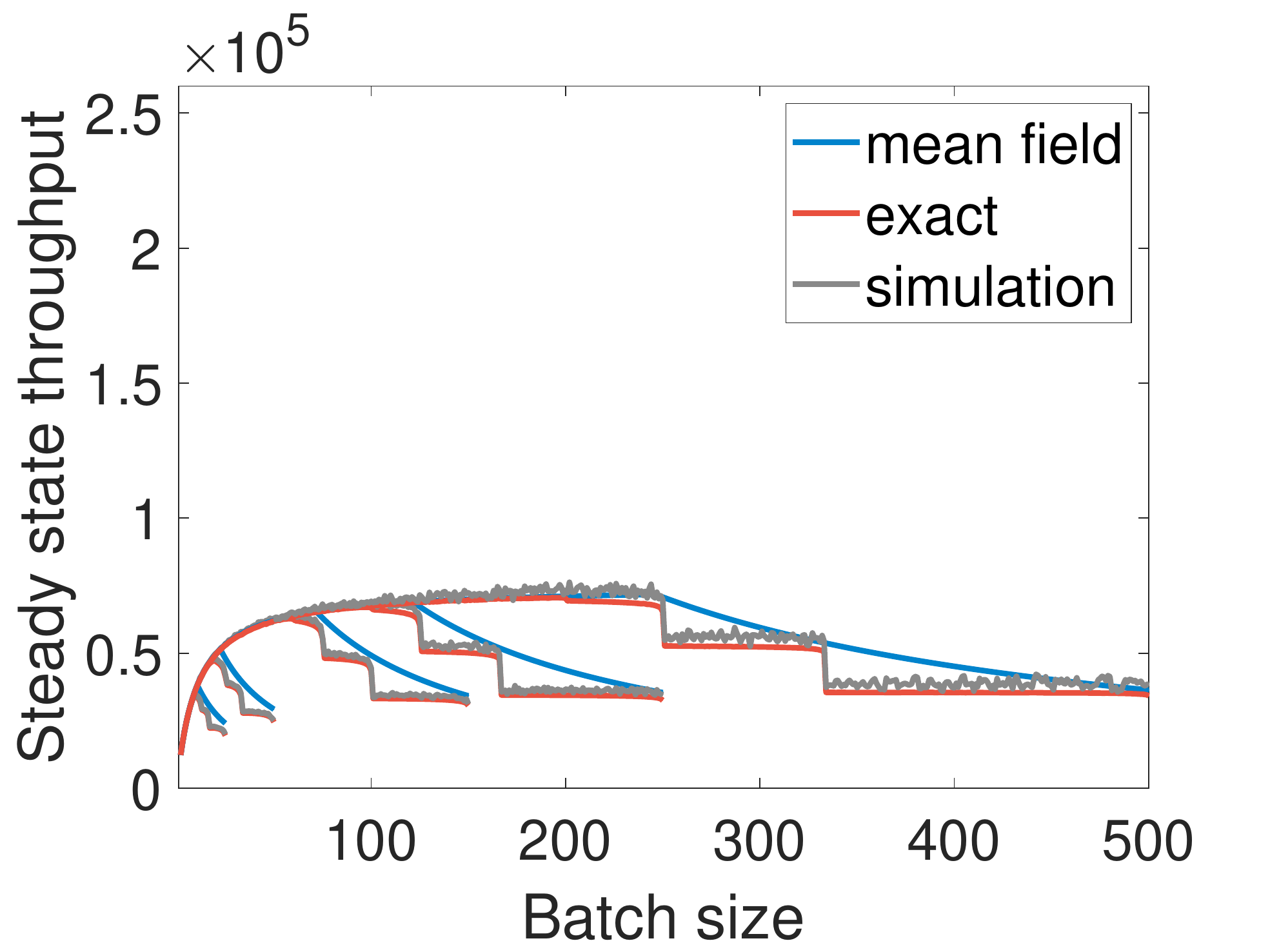}
		\caption{\label{fig:tPutCompare4serv}%
		$m=4$}
	\end{subfigure}
    \hfill
	\begin{subfigure}[t]{0.32\textwidth}
		\centering
		\includegraphics[width=1\textwidth]{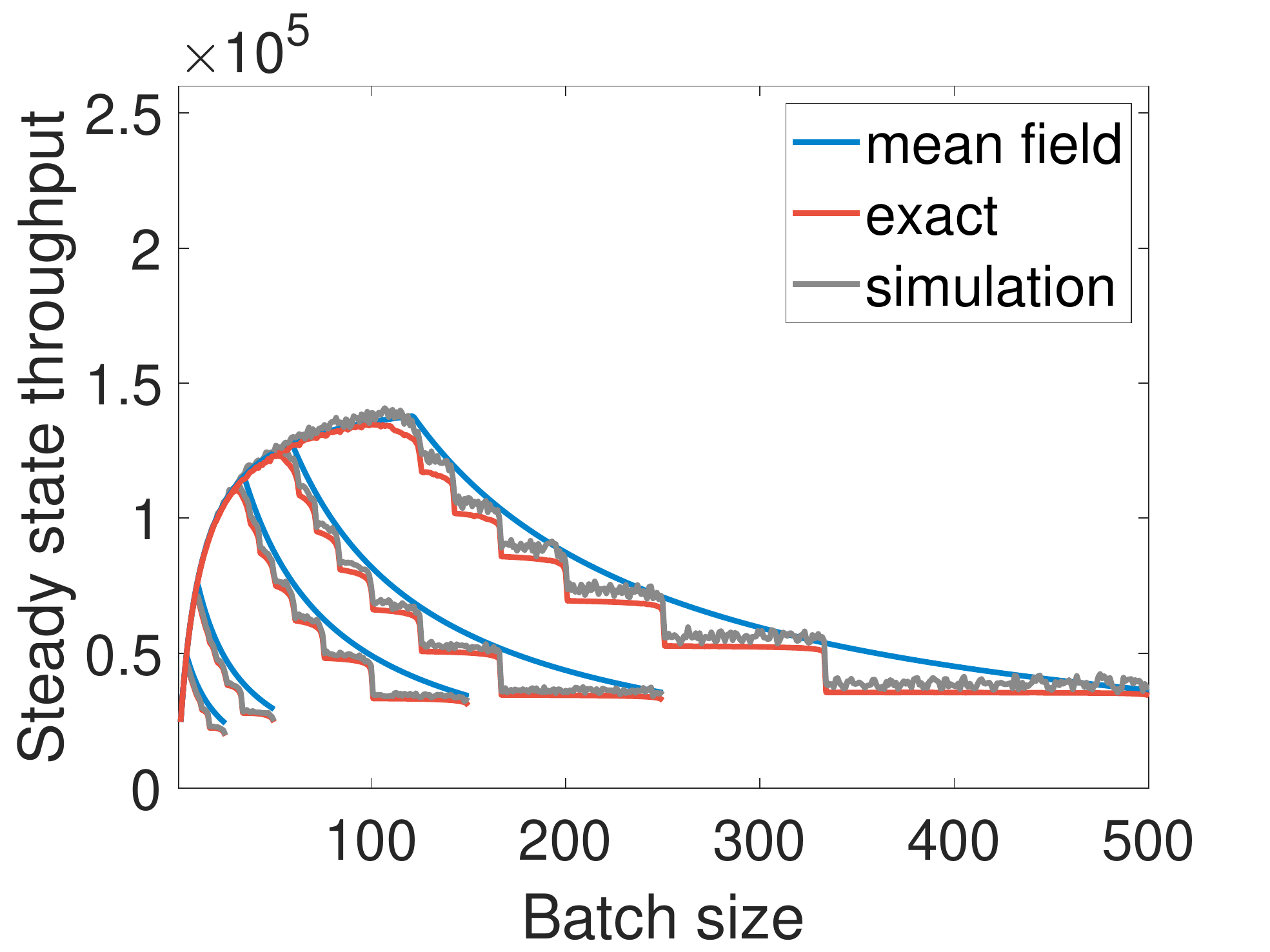}
		\caption{\label{fig:tPutCompare8serv}%
        $m=8$}
	\end{subfigure}
    \hfill
	\begin{subfigure}[t]{0.32\textwidth}
		\centering
		\includegraphics[width=1\textwidth]{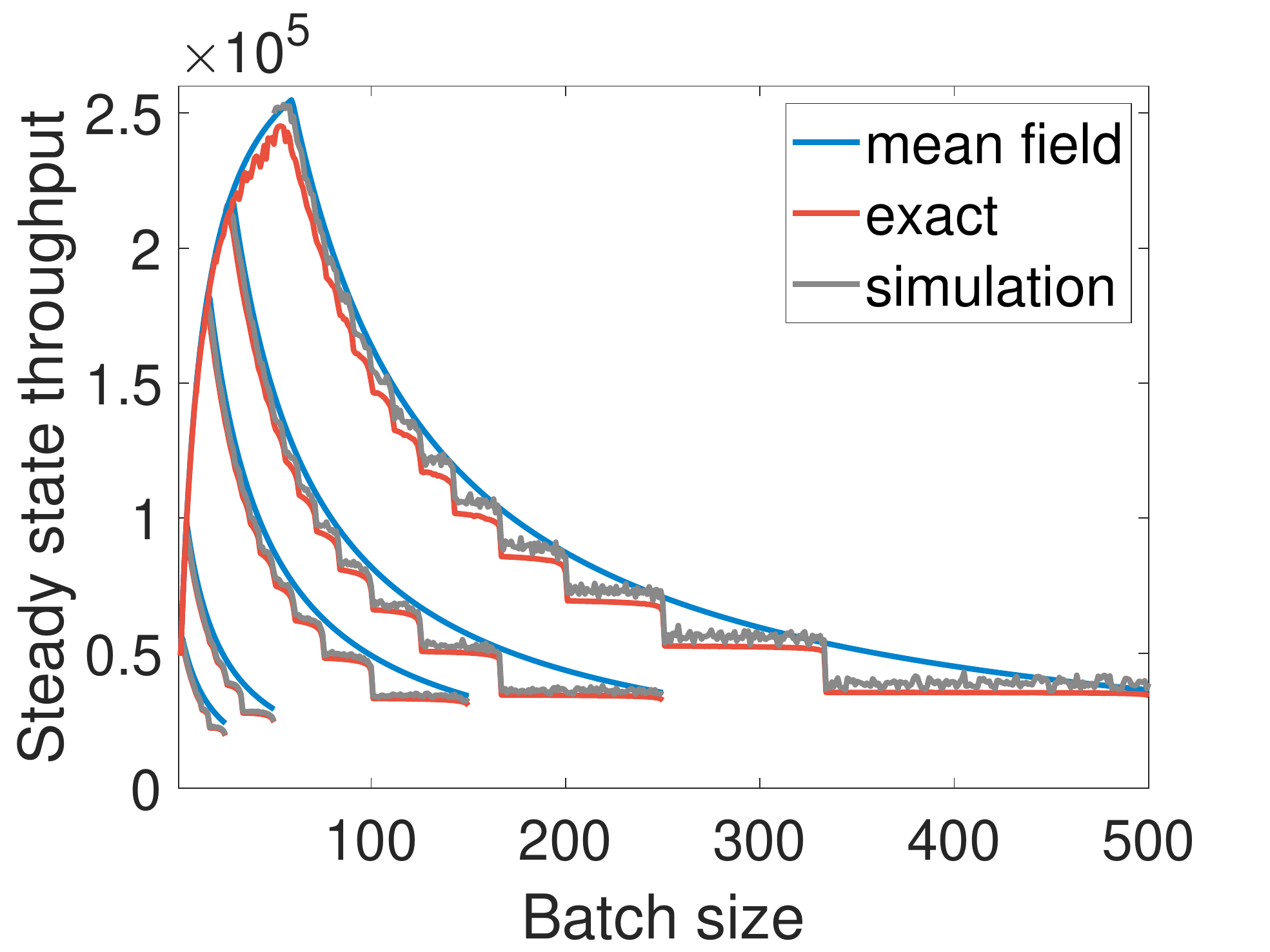}
		\caption{\label{fig:tPutCompare16serv}%
		$m=16$}
	\end{subfigure}
\caption{\label{fig:tPutCompare3}%
Steady state throughput of the system for one job-type for several values of the number of servers $m$; each set of lines corresponds to a value of $n \in [50, 100, 300, 500, 1000]$ in an increasing order (from left to right). Observe that the exact and simulated throughput decreases sharply at points where the number of maximum possible active server drops by one, becoming more apparent for larger batch sizes due to higher relative change. Both the exact and the mean-field model accurately mimic the steady state throughput and accurately capture the optimal batch sizes $k^*$ (from the peak point). }%
\end{figure*}

In Figures~\ref{fig:optTPut-vs-n-1job}-\ref{fig:tPutCompare3}, we compare the steady state throughput for the non-asymptotic/exact model, the mean-field model and simulations, for both the one-job type and the two-job type cases with preemptive priority.  In all figures we vary the number of servers $m$ and obtain the corresponding steady state throughput as a function of the number of clients $n$ or of the batch size $k$. 

\begin{figure*}[t!]
	\centering
	\begin{subfigure}[t]{0.32\textwidth}
		\centering
		\includegraphics[width=1\textwidth]{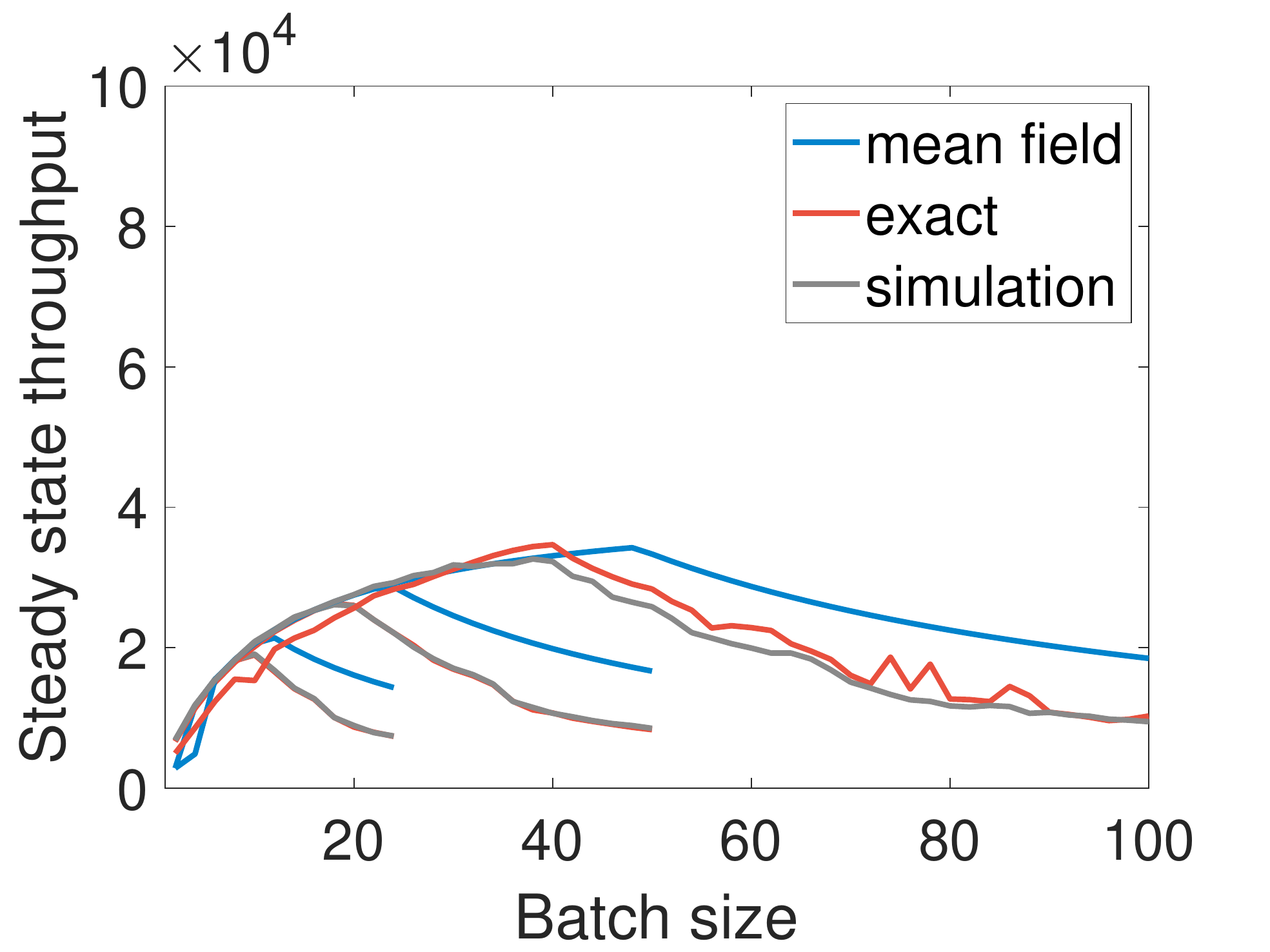}
		\caption{\label{fig:tPutCompare4servPreEmp}%
		$m=4$}
	\end{subfigure}
    \hfill
	\begin{subfigure}[t]{0.32\textwidth}
		\centering
		\includegraphics[width=1\textwidth]{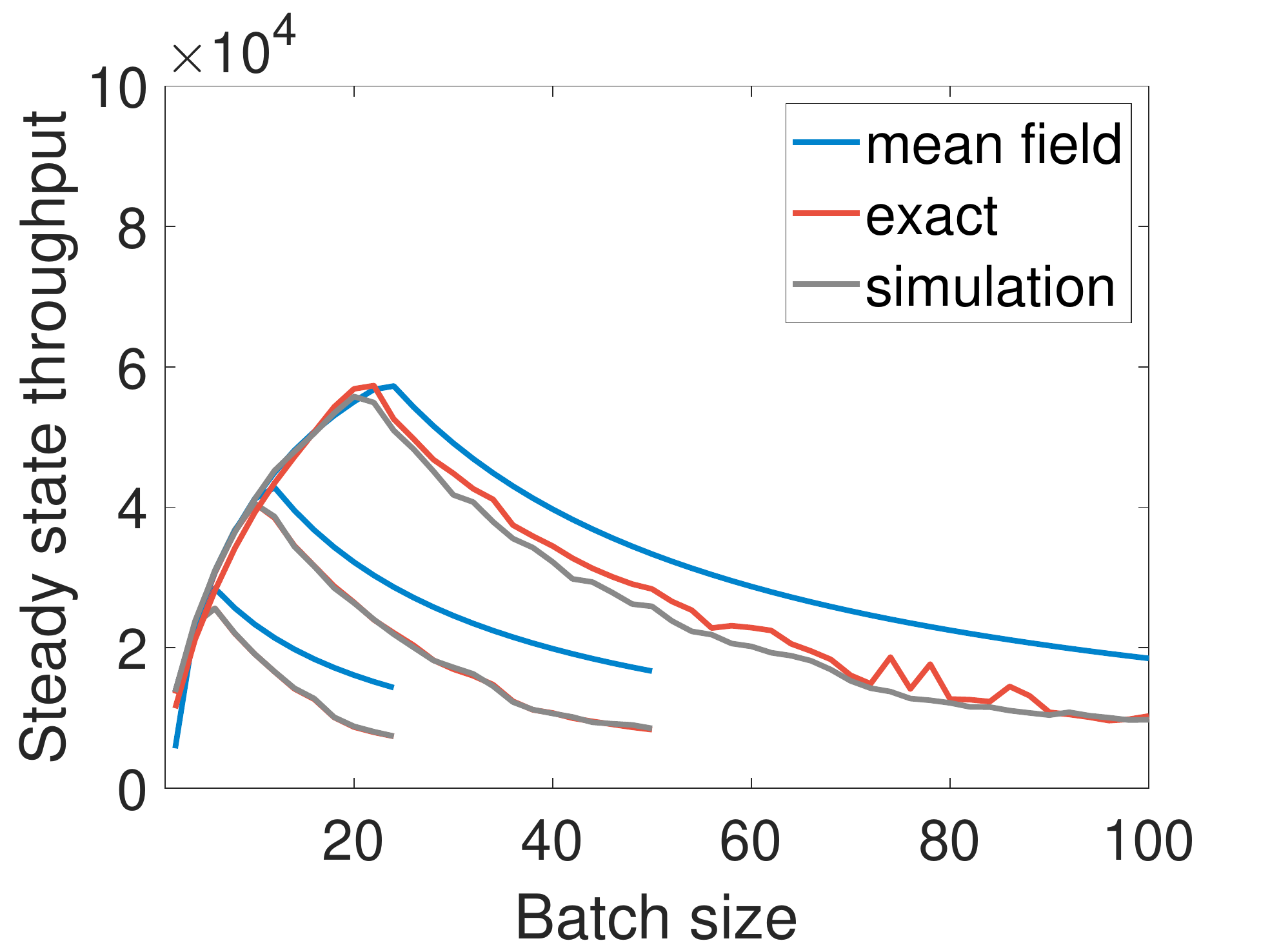}
		\caption{\label{fig:tPutCompare8servPreEmp}%
        $m=8$}
	\end{subfigure}
    \hfill
	\begin{subfigure}[t]{0.32\textwidth}
		\centering
		\includegraphics[width=1\textwidth]{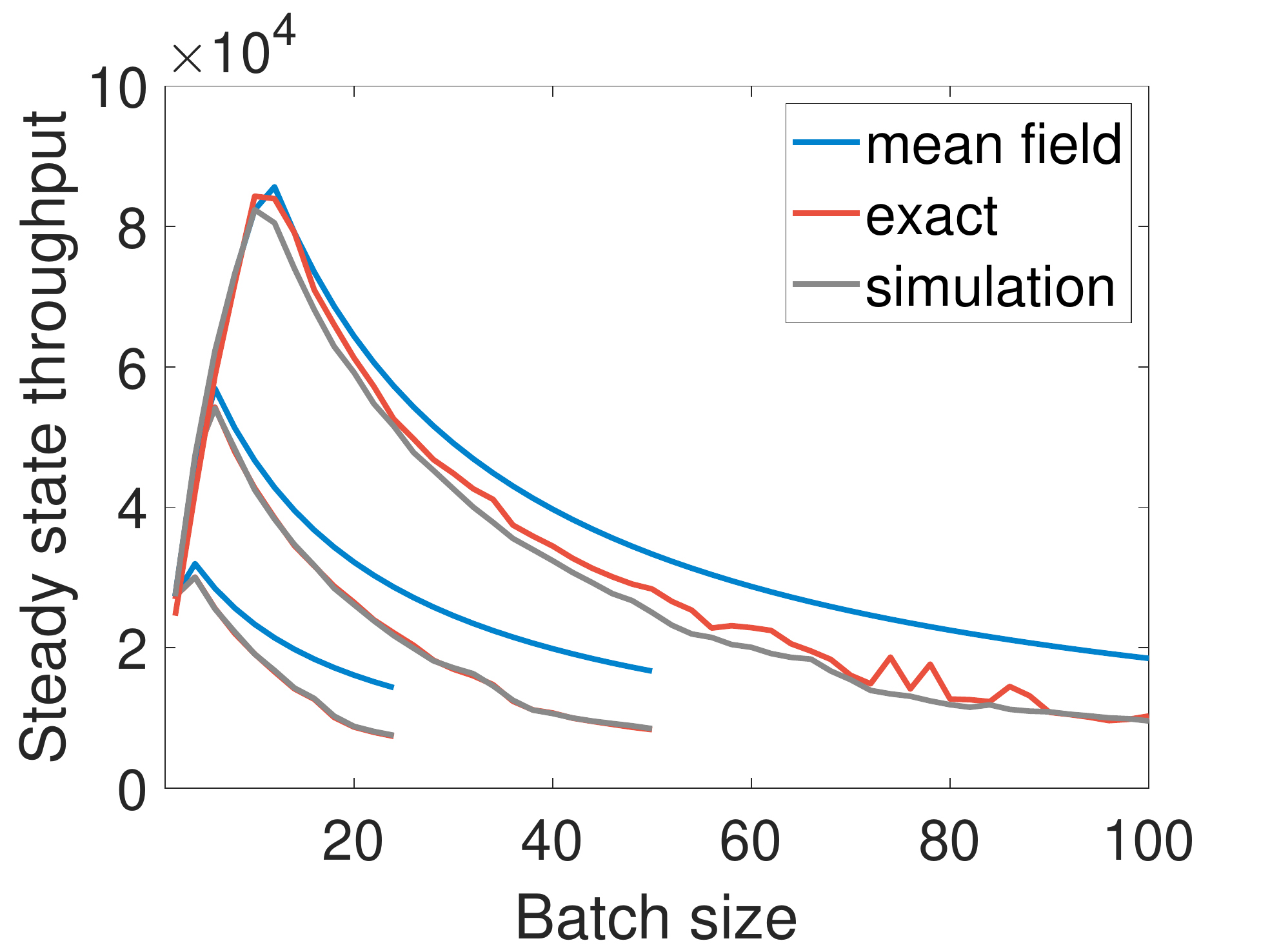}
		\caption{\label{fig:tPutCompare16servPreEmp}%
		$m=16$}
	\end{subfigure}
\caption{\label{fig:tPutCompare3PreEmp}%
Steady state throughput with two job-types and preemptive priority for several values of $m \in [4,8,16]$; each set of lines corresponds to a value of $n \in [50, 100, 200]$ in an increasing order (from left to right).}%
\end{figure*}

\begin{figure}[h!]
	\centering
	\begin{subfigure}[t]{0.48\columnwidth}
		\centering
		\includegraphics[width=1\textwidth]{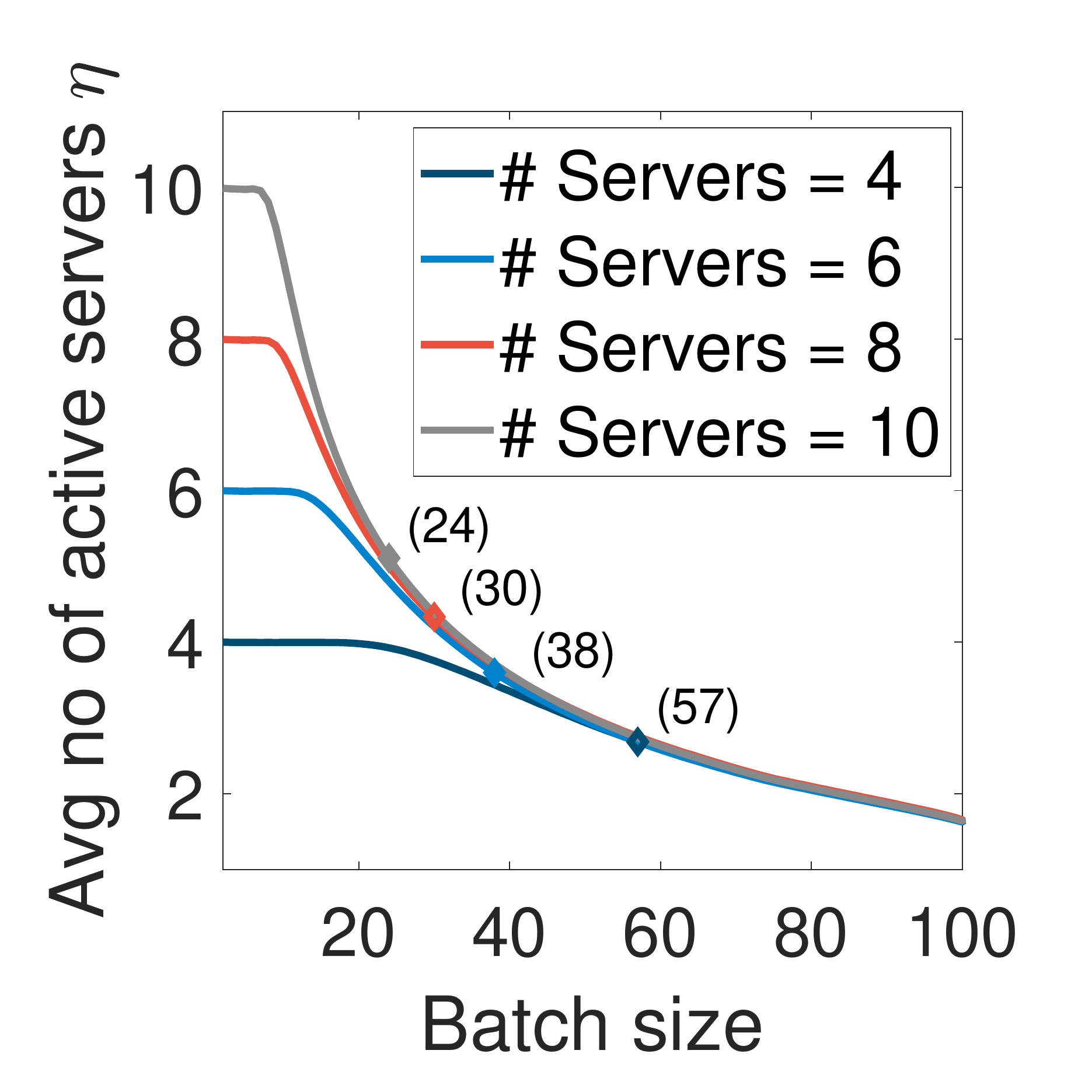}
		\caption{\label{fig:avgActiveServer}%
        Average number of active servers $\eta$ in the steady state  with $300$ clients  attached.}
	\end{subfigure}
	\hfill
	\begin{subfigure}[t]{0.48\columnwidth}
		\centering
		\includegraphics[width=1\textwidth]{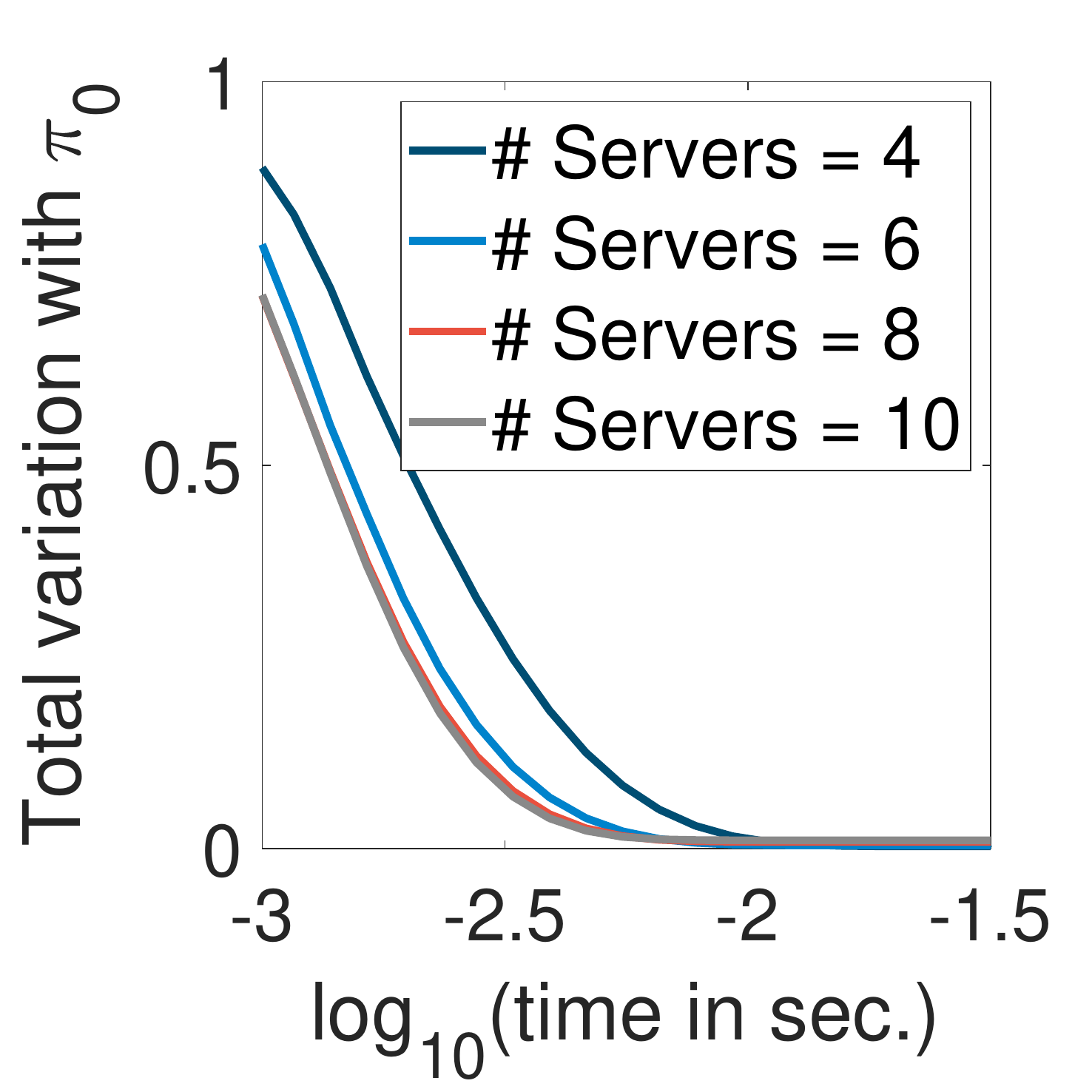}
		\caption{\label{fig:mixing}%
		Total variation distance with steady state distribution $\pmb{\pi_0}$ over time.}
	\end{subfigure}
\caption{\label{fig:others2}%
Steady state and transient characteristics from exact analysis for the system with $n=300$ clients with one job type. Fig.~\ref{fig:others2}(a) shows the average number of active servers $\eta$ in the steady state. The annotated optimal batch sizes show the point until which the speedup compensates for diminishing server utilization. 
Fig.~\ref{fig:others2}(b) shows the total variation distance with the steady state distribution $\pmb{\pi_0}$  for the respective optimal batch sizes over time, i.e., how the marginal distribution of the system states gets reasonably close to the steady state distribution $\pmb{\pi_0}$ within $10$~ms.
}%
\vspace{-10pt}
\end{figure}

In Fig.~\ref{fig:optTPut-vs-n-1job} we show the optimal steady state throughput as a function of the number of clients $n$, for fixed values of the number of servers $m$.
The non-asymptotic/exact model and more interestingly the mean-field model accurately capture the optimal steady state throughput obtained from simulations. The optimal throughput is concave in the number of clients $n$, as it is given by the mean-field analysis as $m k^* \mu(k^*)$ in the limit with $k^*$ from~\eqref{eq:optimal_k_mean_field}.
Similar observations hold in Fig.~\ref{fig:optTPut-vs-n-2jobs} depicting the optimal total steady state throughput for the two job-type case with preemptive priority.

The next set of results in Figs.~\ref{fig:tPutCompare3}-\ref{fig:tPutCompare3PreEmp} concern with the steady state throughput as a function of the batch size $k$. In Fig.~\ref{fig:tPutCompare3} we show how the exact/non-asymptotic model and the mean-field model accurately capture the simulated steady state throughput and provide the optimal batch sizes $k$.
Figure~\ref{fig:tPutCompare3PreEmp} shows the total throughput for the case of two job-types with preemptive priority.

Next, we consider the trade-off between the speedup and the idling of servers. Fig. \ref{fig:avgActiveServer} shows the extent to which the effect of idling is compensated by the batching speedup for a set-up with $n=300$ clients and different number of servers.
For this same set-up, we also look at the convergence rate of the system to the steady state in Fig. \ref{fig:mixing}, but only for the optimal batch size given by the exact analysis. We assume the system starts at the state where all jobs are at the producer/clients station and numerically compute the marginal distribution at regular time intervals. To visualize the distance of the marginal with the steady state distribution of the system, we use the total variation distance as defined in~\cite{LevinPeresWilmer2006}. \looseness=-1 

To conclude this subsection, we note that the mean-field results accurately capture both the optimal steady state throughput and the corresponding optimal batch size $k^*$ of the system.

\subsection{Experimental Evaluation}
In this section, we discuss the performance of our system for experimental evaluations. We start with the description of the system and data collection before comparing our results to actual observation. 
\subsubsection{System Layout} \label{subsec:system_layout}

Here we provide an overview of our system and the Telecom Application Transaction Processing  (TATP) benchmark \cite{neuvonen2009tatp} that is used to retrieve the data for our model.
We run our experiments in a research prototype based on a commercial in-memory database. The database receives a client-request as an SQL string and compiles it to optimized execution plans or extracts such plan from a plan cache, if the string was already compiled for a previous request. Each plan consists of several data-operators, e.g., for accessing tables by index or scanning, or aggregating results, as well as operators for sending the results back to the requesting client. \looseness = -1

\begin{figure}[t]
	\centering
	\includegraphics[width=0.98\linewidth]{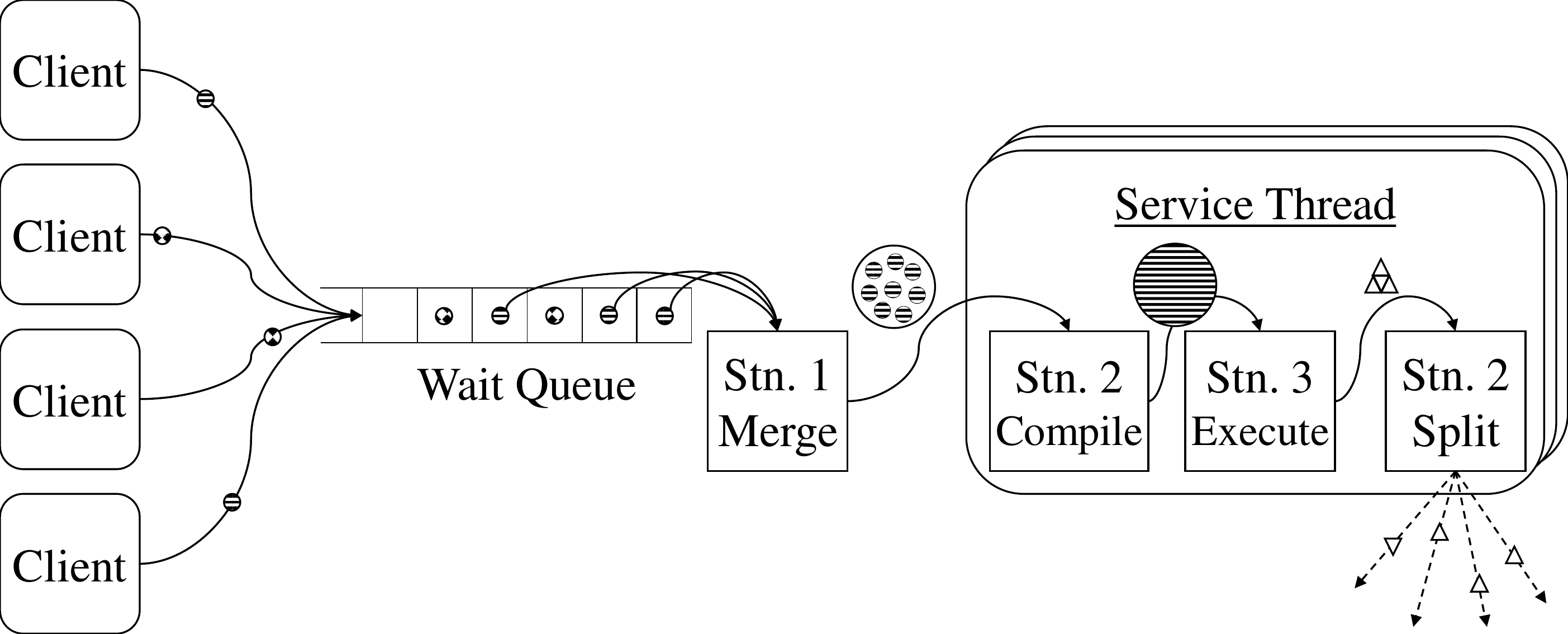}
    \caption{Query Batching in the Database System. Requests of the same SQL string are merged and executed as a batch.
    }
    \label{fig:system_layout}
    \vspace{-5pt}
\end{figure}

Fig.~\ref{fig:system_layout} shows that incoming requests are not executed instantly but rather wait in a queue, until the number of waiting requests reaches a certain threshold (i.e., the batch size).
Once this event occurs, the number of requests to grab from the waiting queue is determined, we extract that amount of requests, preferring the write jobs and create one SQL string from the requests.
The service thread then compiles and executes the merged SQL string, which produces a shared result. Finally, the service thread splits the shared result to return to each client its individual result.

Service threads execute three tasks on a merged batch taken from the waiting queue:
\begin{enumerate*}
    \item\label{enum:compile} compilation,
    \item\label{enum:exec} execution, and
    \item\label{enum:split} splitting the results.
\end{enumerate*}
For merging, we need to execute some string operations to create the merged SQL string. The processing time of this step depends on the number of requests extracted from the waiting queue. In comparison, step (\ref{enum:compile}) first looks up the cache, whether that SQL string was already compiled and only if this is not the case, it compiles the string itself. This is a crucial step, because compiling a string into an executable plan is a time consuming task.
The execution of a batch in step (\ref{enum:exec}) heavily depends on the table format (row-store or column-store \cite{turner1979columnstore}) and whether an index exists on the filtered column or the column needs to be scanned. And finally, in the last step (\ref{enum:split}), the service thread scans the shared result for each client that belongs to the batch and sends back the matching rows.
\subsubsection{Experiment and Data Description}
\label{sec:expDesc}
For our experiments, we focus on two transactions of the TATP benchmark \cite{neuvonen2009tatp}, a well known Online Transactional Processing (OLTP) benchmark for databases. The two transactions used are GET\_SUBSCRIBER\_DATA, consisting of one read operation and the DELETE\_CALL\_FORWARDING, consisting of one read and one write, namely a delete operation. Each operation is expressed as an SQL string, which is sent to the database and processed on the server side, as described earlier. Each of the reading and writing operations access only one row of exactly one table to read or delete from and are usually processed in less than \SI{1}{\milli\second}. We adjust the DELETE\_CALL\_FORWARDING transaction in such a way that it submits a single read operation in 80\% of all cases and a delete operation in the remaining 20\%.

We run our experiments on a base table size of $10^4$ rows with a varying number of clients.
The database and the clients run on different sockets of the same server with SUSE Linux Enterprise Server 12 SP1 (kernel: 4.1.36-44-default), having \SI{512}{\giga\byte} of main memory, four sockets with \SI{10}{cores} each and no hyperthreading.
The server runs on Intel(R) Xeon(R) CPU E7-4870, with a speed of \SI{2.4}{\giga\hertz} and a cache size of \SI{30}{M\byte}.

Internally, we keep track of the job arrival and retrieval times from the queue, as well as the execution time of its batch.
This sums up the data retrieved from the experiments and used for creating and validating the model.

\subsubsection{Fitting the Experimental Data} \label{subsec:dataFitting}

In the following, we employ standard optimal experiment design techniques to characterize the service distributions for all batch sizes, while letting the batch-processing system run only for some selected batch sizes. To this end, we estimate the batching speedup and characterize the corresponding service distributions. For the sake of brevity, we describe the estimation process for only one job type; the two job-type case proceeds similarly.

First, we express the batching speedup through the function $g:\setOfNaturals \mapsto \setOfPositiveReals$ where $g(k) = 1/\mu(k)$.
To avoid triviality, we assume sub-additivity, i.e.,  $g(k_1+k_2) \le g(k_1)+g(k_2)$. In the experimental evaluation, we consider the best fit of the empirical data to have one of the following speedup forms:
\begin{itemize}
    \item $g_1(k) = ak+b$ with $a<1$
    \item $g_2(k) = \gamma k^\alpha$ with $\alpha<1$
    \item $g_3(k) = c \log k+d$ with $c<1$.
\end{itemize}
Each speedup function is characterized by some parameters which are estimated by fitting the mean service times for different batch sizes.

To estimate the speedup function in the given commercial database system we calculate a set of batch sizes which minimize the estimation error. Our approach is based on a linear regression where we transform the speedup function into a linear combination  of weights $\mathbf{w}$ and feature vectors $\pmb{\phi}(k)$. Assuming a Gaussian distribution on the error of the responses of this model, i.e., the mean service times, the standard linear model can be used and hence the ordinary least square (OLS) regression estimate of the regression weights can be found. For the experiment design on the batch-processing system, i.e., deciding on the set $A$ containing which batch sizes to run for the subsequent fitting, we employ a D-optimal design \cite{pukelsheim1993optimal} to minimize the log determinant of the covariance matrix of the OLS estimator.  The size of the subset $A$ is usually set in accordance with time and cost considerations. We solve this integer optimization problem numerically after relaxation using the CVX package \cite{grant2008cvx}.

Finally, we denote the set of sample service times corresponding to the batch size $k \in A$ as $S_k$, and the respective mean service times as $\mathbf{E}[Y(k)]$, and find the speedup function $g$ minimizing the corresponding OLS estimation error, i.e., $g = g_m$ where $m = \argmin_{i} e_i$ and  $e_i = \sum_{k \in A} \big( g_i(k|\hat{\mathbf{\theta_i}})-\mathbf{E}[Y(k)] \big)^2$. Here, we express the parameter space corresponding to the parameter vector $\mathbf{\theta_i}$ of the speedup function $g_i$ as $\mathbf{\Theta_i}$, and adopt an OLS approach to estimate $\mathbf{\theta_i}$ through $ \hat{\mathbf{\theta_i}} = \argmin_{\theta_i \in \Theta_i} \sum_{k \in A} \big( g_i(k)-\mathbf{E}[Y(k)] \big)^2$.\looseness=-1

\begin{figure}[t!]
	\centering
	\begin{subfigure}[t]{0.48\columnwidth}
		\centering
		\includegraphics[width=1\textwidth]{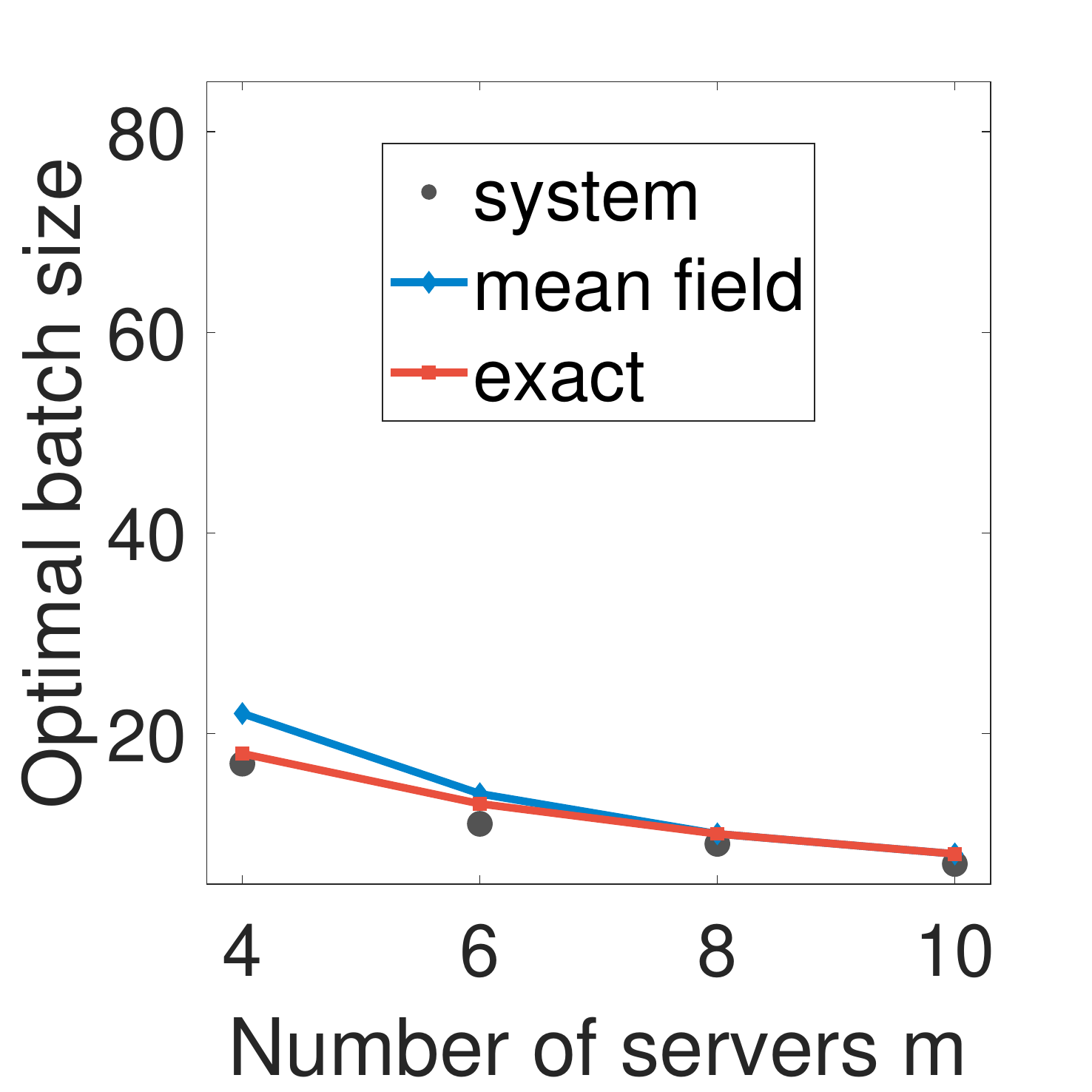}
		\caption{\label{fig:serverWiseOptBatch100}%
		$n=100$}
	\end{subfigure}
    \hfill
	\begin{subfigure}[t]{0.48\columnwidth}
		\centering
		\includegraphics[width=1\textwidth]{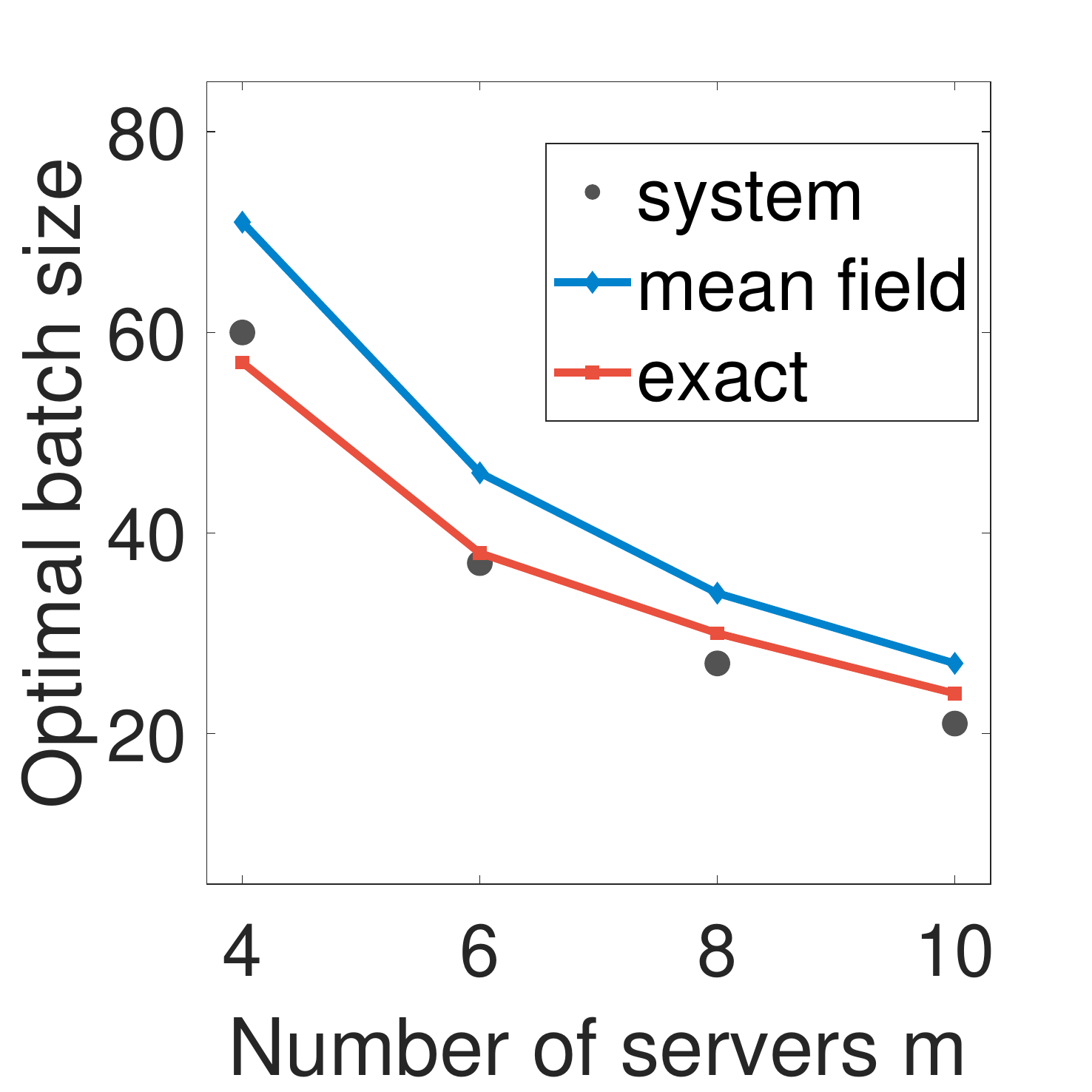}
		\caption{\label{fig:serverWiseOptBatch300}%
        $n=300$}
	\end{subfigure}
\caption{\label{fig:others1}%
Experimental evaluation: Comparison of the observed optimal batch sizes $k^*$ and the model estimates with increasing number of servers. The system receives only one job type, i.e., \textit{read} jobs, and the comparison is done for a varying number of clients. As expected the optimal batch size decreases with increasing number of servers due to server idling.}%
\vspace{-7pt}
\end{figure}

\begin{figure}[t!]
	\centering
	\begin{subfigure}[t]{0.48\columnwidth}
		\centering
		\includegraphics[width=1\textwidth]{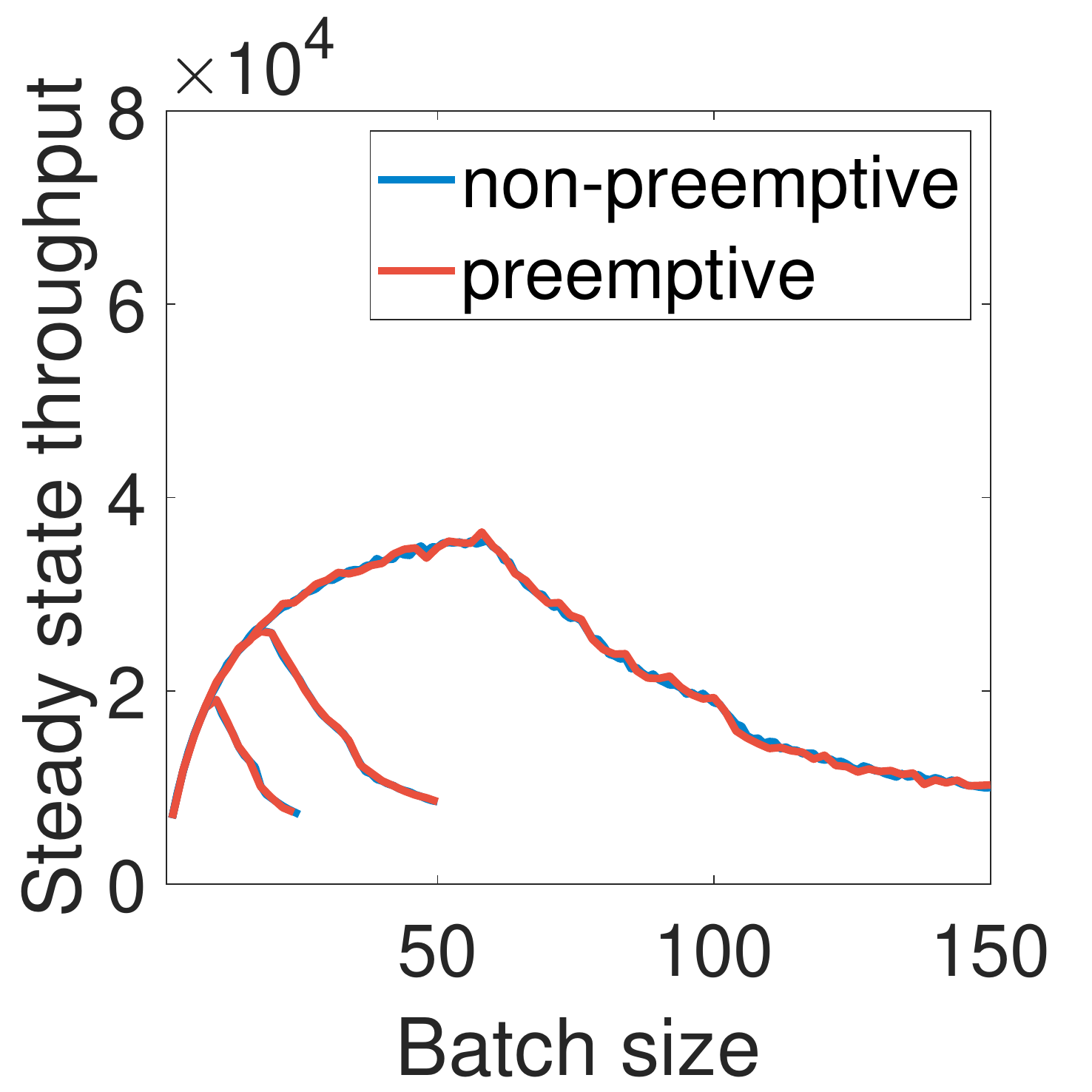}
		\caption{\label{fig:preNpCompare4}%
		$m=4$}
	\end{subfigure}
    \hfill
	\begin{subfigure}[t]{0.48\columnwidth}
		\centering
		\includegraphics[width=1\textwidth]{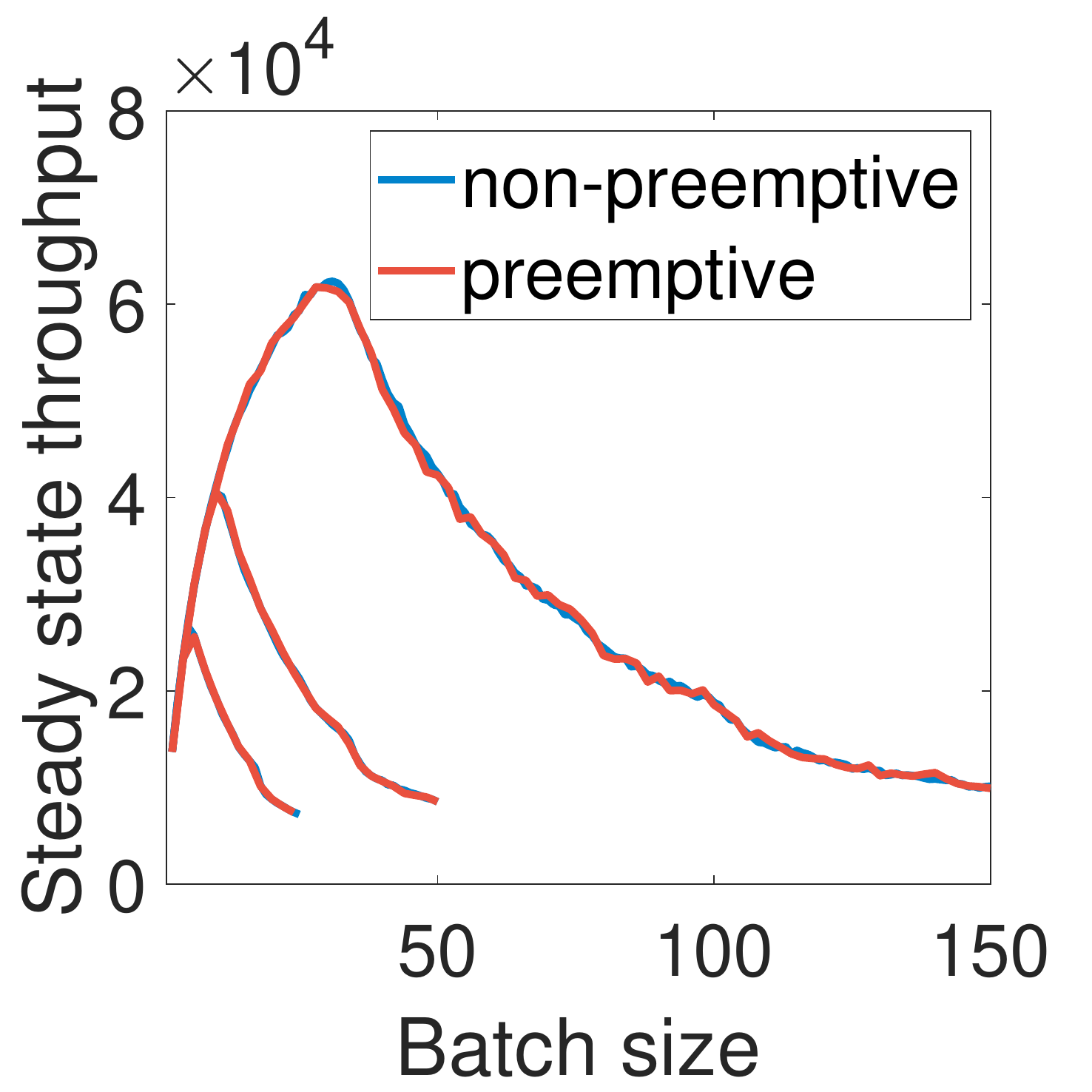}
		\caption{\label{fig:preNpCompare8}%
        $m=8$}
	\end{subfigure}
\caption{\label{fig:preNpCompare}%
Equivalence of preemptive and non-preemptive priority in terms of the steady state throughput for a simulated system with two job-types; each set of lines corresponds to a value of $n \in [50, 100, 300]$ in an increasing order (from left to right).}%
\vspace{-10pt}
\end{figure}

\subsubsection{Evaluation}
For the experimental evaluation we set a measurement budget for the fitting and parameter estimation, i.e., we estimate the service times and the speedup based on measurement runs for only $\sim5\%$ of all possible batch sizes. Using the optimal experimental design approach from the previous section we calculate the set of batch sizes to be measured $A$ for $n\in\{100,300\}$ clients. 
For each $n$ we estimate the mean batching and service times for each batch size $k \in A$ from independent runs. 
The mean service times for batch sizes $k \in A$ are then used to estimate the speedup.
\looseness=-1
Equipped with the estimated service and batching rates, we populate the intensity matrix $Q$ using~\eqref{eq:jumpRates} and subsequently solve for the steady state distribution. We further calculate the steady state throughput using \eqref{eq:steadyThroughput} and obtain the corresponding optimal batch size. 
We repeat the same process for a varying number  of servers $m$ and for a varying number of clients $n$ up to $300$.
Note that the database prototype at hand has at most $m=10$ available servers. 

In addition, we run an exhaustive experiment for all possible batch sizes to find the empirical optimum for the set-up with a varying number of servers and clients for the sake of completeness. Fig.~\ref{fig:others1} shows a comparison of the modelled and observed optimal batch sizes $k^*$ for an increasing number of servers and  different number of clients $n$. We observe that our models are accurate. Both the non-asymptotic/exact model as well as the mean-field model  capture the decline in the optimal batch size with an increasing number of servers $m$. 

\begin{figure}[t]
	\centering
	\includegraphics[width=0.98\linewidth]{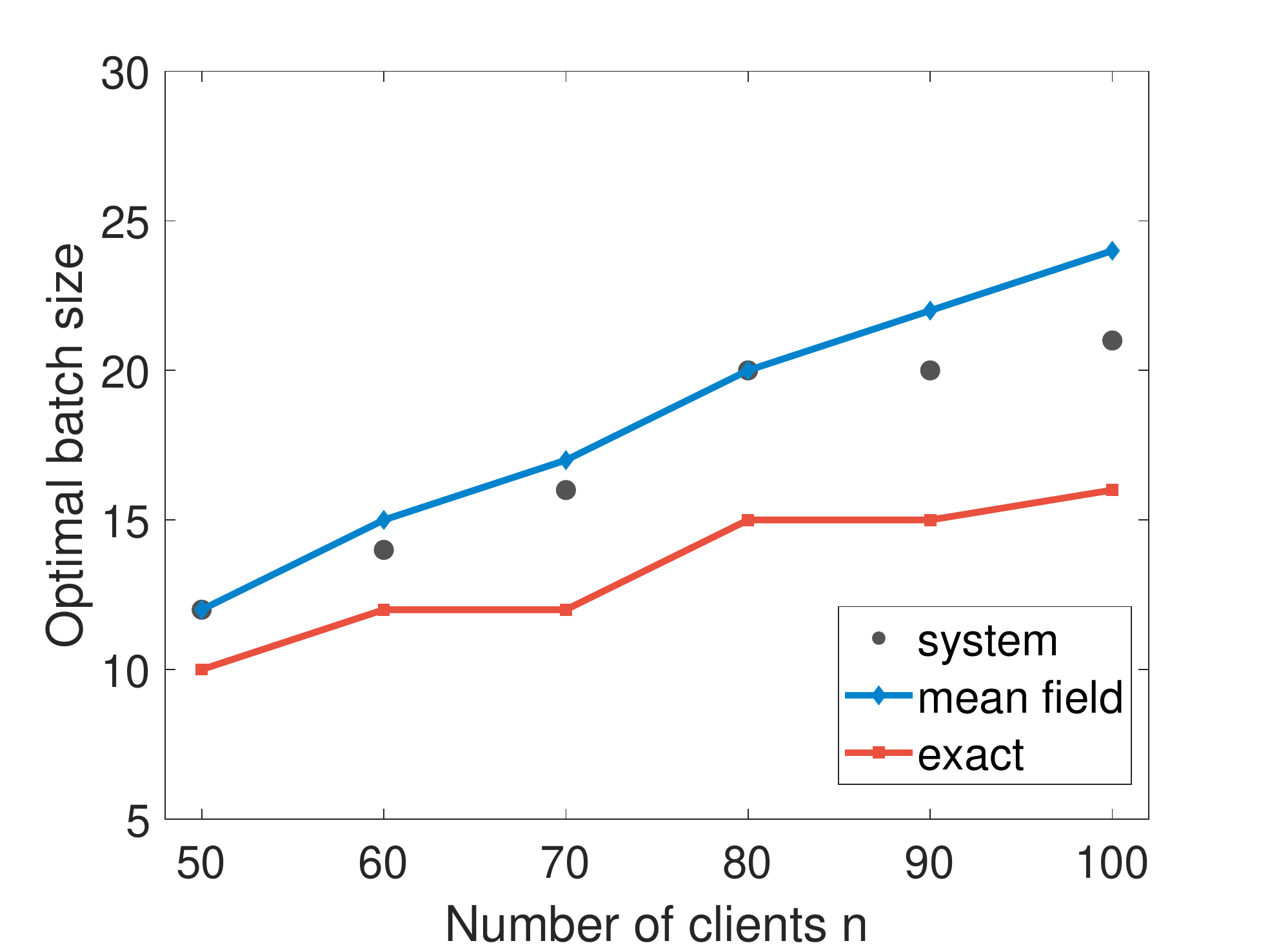}
    \caption{ Optimal batch sizes for \textit{read} and \textit{write} job types with $m=4$ servers and a varying number of clients $n$. The system implements non-preemptive priority of \texttt{write} jobs over \texttt{read} jobs. For mean field analysis, optima are approximated by the preemptive model in Sect. ~\ref{subsec:mFTwoJobs} whereas the exact model follows the workflow in  Sect. ~\ref{sec:app-A}. As expected, modelled and observed optima rise in close proximity. 
    }
    \label{fig:compareRW}
    \vspace{-10pt}
\end{figure}%

We also conduct experiments where the submitted jobs can be of two types: \textit{read} or \textit{write}. A new request can be a \textit{read} query with probability $0.8$ and a \textit{write} query with probability $0.2$. Further, the \textit{write} jobs have priority over the \textit{read} jobs. The prototype system provides a non-preemptive priority to the \textit{write} jobs; however, the difference in the system throughput diminishes in the stationary asymptotic regime, as illustrated through simulations in Fig. ~\ref{fig:preNpCompare}. In Fig.~\ref{fig:compareRW} we compare the modelled and the actual optimal batch sizes in the system for the two job case for a varying number of clients and observe a reasonably close match. 
The contributed mean-field model is seen to capture the system behavior very well. 
\section{Conclusion}\label{sec:conclusion}

In this work, we optimize the throughput of closed data-processing systems that process incoming jobs in batches. Through modelling the system as a closed queueing network, where batches observe a sub-additive speedup in execution, we obtain the optimal throughput as a function of the batch size for $n$ clients and $m$ servers. The considered system resembles standard database systems where clients wait for the result of an input query to generate the next one. 
We contribute a mean-field model that captures the system throughput in the asymptotic regime and show that the analytical results accurately provide the optimal throughput, as well as, the corresponding optimal batch size in simulation, as well as, for a prototype of a large commercial system. 

\vspace{-2pt}
\bibliographystyle{ACM-Reference-Format}
\bibliography{sample-base}
\newpage
\appendix
\section{Appendix}
\label{sec:appendix}

\subsection{Irreducibility of the Closed Queueing System} \label{sec:app-0}

\begin{prop}
The Markov chain describing the queueing system in Sect. ~\ref{sec:modelSingleJobType} is irreducible.
\end{prop}%
\vspace{-10pt}
\begin{proof}
It is sufficient to show that the states $(n,0,0)$ and $(x,y,zk)$ communicate. To show that $(x,y,zk)$ can be reached from $(n,0,0)$ in finite steps with positive probability, we show that each of the intermediate states in the following can be reached in finite steps with positive probability:
\begin{align*}
    &(n,0,0) \xrightarrow{} (n-k,k,0) \xrightarrow{} (n-y-zk,y+zk,0)\\ &\xrightarrow{} (n-y-zk,y+(l-1)k,k) \xrightarrow{} (n-y-zk,y,zk).
\end{align*}
Starting from $(n,0,0)$, $(n-k,k,0)$ is reached in $k$ steps with probability 1. This is due to the fact that there can not be any batching unless there are at least $k$ jobs at the batching station. Further, $(n-y-zk,y+zk,0)$ is reached in another $y+(l-1)k$ steps where the $r$th step has probability $p_r = P[X_r<Y]$. Here, $X_r$ is an exponential variable with mean $1/((n-k-r+1) \lambda)$ and $Y$ is another independent exponential variable with mean $1/M$. Each of these steps corresponds to the outcome that the producer sends a job to the batcher before it could form a batch. Again, $(n-y-zk,y+(l-1)k,k)$ is reached from $(n-y-zk,y+zk,0)$ in a single step with probability $P[Y<X)]$ where $X$ is another independent exponential variable with mean $1/((n-y-zk) \lambda)$. That is, a batch is formed by the batcher before the dispatcher could send a new job. Finally, $(n-y-zk,y,zk)$ is reached in another $(l-1)$ step where the $r$th step has probability $P[Y<\min(X,Z_r)]$, $Z_r$ being an exponential variable with mean $1/(\min(n,r) \mu(k))$. Each step describes the event that the batching station merges a batch before either the dispatcher could send a new job or the sever could finish serving a batch.
Similarly, we can show that starting from $(x,y,zk)$, there exists a way to reach $(n,0,0)$ in finite steps with positive probability, completing the proof.
\end{proof}  
Since the Markov chain describing the states of the queueing system in Sect. \ref{sec:modelSingleJobType} is finite and irreducible, it is positive recurrent as well. Thus, there exists a unique steady state distribution for this chain that is obtainable by solving the equation $\pmb{\pi} \cdot \mathbf{Q} = 0$. Similarly, we can argue about the existence and uniqueness of the steady state distribution for the system described in Sect. \ref{sec:modelMultJobType}.

\subsection{System with Two Job Types and Non-preemptive Priority} \label{sec:app-A}
Unlike the case with preemptive priority in Sect. ~\ref{sec:modelMultJobTypeBig}, the case with non-preemption requires the number of type $1$ jobs in service explicitly. The system can be uniquely described by the tuple $(x,y_1,y_2,u_1 k_1,v_1 k_1)$ where $x$ is the number of active clients, $y_{\iota}$ is the number of type $\iota$ jobs not yet batched, $u_1$ is the number of type $1$ batches waiting in the queue and $v_1$ is the number of type $1$ batches in service and $s = (x,y_1,y_2,u_1 k_1,v_1 k_1)$ belongs to the state space 
\vspace{-10pt}
\begin{align*}
    \mc{S}=\cbrac{(x_1,x_2,x_3,x_4,x_5):\in \mb{Z}_+^5: \mathbf{x}.\mathbf{1} \le n, k|x_4, k|x_5}. 
\end{align*}
Here $\mathbf{1}$ denotes the column vector of ones whose size is implied from the context. Note that the number of type $2$ batches is given by $z_2 = (n-s.\mathbf{1})/k_2$ out of which $v_2 = \min(m-v_1,z_2)$ many are in service. The system evolves as CTMC and the jump rates are:
\begin{align}
\label{eq:jumpRatesNonPreemptive}
    s &\xrightarrow[\text{}]{\lambda x p} s-\mathbf{e_1}+\mathbf{e_2}, \thinspace x>0 \nonumber \\
    &\xrightarrow[\text{}]{\lambda x (1-p)} s-\mathbf{e_1}+\mathbf{e_3}, \thinspace x>0 \nonumber \\
    &\xrightarrow[\text{}]{M_1(k_1)\floor{y_1/k_1}} s-k_1\mathbf{e_2}+k_1\mathbf{e_4}, \thinspace y_1 \ge k_1, v=m \nonumber \\
    &\xrightarrow[\text{}]{M_1(k_1)\floor{y_1/k_1}} s-k_1\mathbf{e_2}+k_1\mathbf{e_5}, \thinspace y_1 \ge k_1, v<m \nonumber \\
    &\xrightarrow[\text{}]{M_2(k_2)\floor{y_2/k_2}} s-k_2\mathbf{e_2}, \thinspace y_2 \ge k_2 \nonumber \\
    &\xrightarrow[\text{}]{v_1\mu_1(k_1)} s+k_1\mathbf{e_1}-k_1\mathbf{e_5}, \thinspace v_1 \ge 1, u_1=0 \nonumber\\
    &\xrightarrow[\text{}]{v_1\mu_1(k_1)} s+k_1\mathbf{e_1}-k_1\mathbf{e_4}, \thinspace v_1 \ge 1, u_1 \ge 1 \nonumber\\
    &\xrightarrow[\text{}]{v_2\mu_2(k_2)} s+k_2\mathbf{e_1}, \thinspace v_2 \ge 1, u_1 = 0 \nonumber\\
    &\xrightarrow[\text{}]{v_2\mu_2(k_2)} s+k_2\mathbf{e_1}-k_1\mathbf{e_4}+k_1\mathbf{e_5}, \thinspace v_2 \ge 1, u_1 \ge 1,
\end{align}
where $s=(x,y_1,y_2,z_1 k_1)$ and $v=v_1+v_2$ denotes the total number of busy servers. The jump rates to any other state is zero. Similar to Sect. ~\ref{sec:modelMultJobType}, we can solve $\pmb{\pi} \cdot \mathbf{Q} = 0$ to get the steady state distribution $\pmb{\pi}_0$, derive the optimal throughput and find the optimal batch size $k^*$ for maximum throughput.

\subsection{Extension to Multiple Job Types with Preemptive Priority} \label{sec:manyJobManyService}
Let us recall the framework described in Sect. \ref{subsec:mFTwoJobs} and consider the case that there are $r$ types of jobs in the system with job of type $i_1$ having preemptive priority over type $i_2$ whenever $i_1<i_2$. We suppose that each client produces a job of type $i$ with probability $p_i$ where $\sum p_i = 1$.
Further, we assume batches go through $d$ levels of service before being unbatched and finally individual responses are sent back to the clients. The workflow of the system requires that after each level of service, batches wait in a common queue if all servers of the next stage\footnote{we use level/stage interchangeably} are busy. Let $k_i$ denote the batch size for job type $i$ for all stages, $m_j$ denote the total number of servers at stage $j$ and $\mu_{ij}(k_i)$ denote the service rate of type $i$ at level $j$ for batch size $k_i$. We will suppress the argument for $\mu_{ij}$ when the dependence is clear. If $X_{ij}^{(n)}(t)$ denotes the number of type $i$ jobs that are waiting for or are at $j$-th level of service, we see that $(X_{ij}^{(n)}(t), 1\le i \le r, 1 \le j \le d, t\geq 0)$ is Markov on state space $\mc{S}=\cbrac{(x_{ij})\in \mb{Z}_+^{rd}: \sum_{i=1}^r \sum_{j=1}^d x_{i,j} \leq n}$. Note that $X_{1j}$ includes unbatched jobs of type $j$ as well whereas $X_{ij}$, $i>1$, only comprises of batches. We consider the corresponding scaled process $w^{(n)}(t)=(w_{ij}^{(n)}(t))$,$w_{ij}^{(n)}(t)=X_{ij}^{(n)}(t)/n$, $1\le i \le r, 1 \le j \le d$. We have
\begin{align}\label{eq:diffMultServMultType}
    \dot w_{11} &= \lambda p_1 \brac{1-\sum_{a,b} w_{ab}}- \mu_{11} w_{11}, \\ \nonumber
    \dot w_{i1} &= \lambda p_i \brac{1-\sum_{a,b} w_{ab}}- \mu_{i1} \min \brac{w_{i1},\max \brac{0,\alpha_1-\sum_{l<i}\frac{w_{l1}}{k_l}}k_i },\\ \nonumber
    \dot w_{1j} &= \mu_{1(j-1)} w_{1(j-1)}- \mu_{1j} w_{1j}, \\ \nonumber
    \dot w_{ij} &=  \mu_{i(j-1)} \min \brac{w_{i(j-1)},\max \brac{0,\alpha_{j-1}-\sum_{l<i}\frac{w_{l(j-1)}}{k_l}}k_i } \\ \nonumber
    &- \mu_{ij} \min \brac{w_{ij},\max \brac{0,\alpha_j-\sum_{l<i}\frac{w_{lj}}{k_l}}k_i }, 2\le i \le r, 2 \le j \le d,
\end{align}
where $m_j/n \to \alpha_j$ and $\sum_{i,j} w_{ij} \le 1$. We use the shorthand notation $\frac{d\mathbf{w}}{dt}  = \mathbf{F}(\mathbf{w})$ for \eqref{eq:diffMultServMultType}. We notice that $\mathbf{F}$ is Lipschitz continuous which follows from arguments identical to part (i) of Thm. \ref{thm:meanfield_onetype}. Also, the stationary measure $\pi_w^{(n)}$ is tight
as it is defined on the compact space $[0,1]^{rd}$. We observe that the dynamical system given by \eqref{eq:diffMultServMultType} is piecewise linear and we investigate global attraction to the fixed point when $k_i=k$ and there is only one level of service.

\begin{theorem}\label{thm:glbalAttractionMultJob}
Consider the system in \eqref{eq:diffMultServMultType} when $k_i=k$, $1\le i \le r$ and $d=1$. The system is globally attractive to  
\begin{align*}
    \mathbf{w}^*=\begin{cases}
                        \mathbf{A}^{-1} \mathbf{c_a}, \quad \text{if } \langle \mathbf{A}^{-1} \mathbf{c_a}, 1 \rangle < k \alpha\\
                        \mathbf{B}^{-1} \mathbf{c_b}, \quad \text{otherwise},
                    \end{cases}
\end{align*}
where 
\begin{align*}
    \mathbf{A} &= \begin{bmatrix}
        -\mu_1- \lambda p_1 & - \lambda p_1 & \dots & - \lambda p_1\\
        -\mu_2 & -\mu_2- \lambda p_2 & \dots & - \lambda p_2\\
        \dots & \dots & \dots & \dots \\
        -\mu_r & -\mu_r & \dots & -\mu_r- \lambda p_r
            \end{bmatrix},\\
    \mathbf{B} &= \begin{bmatrix}
        -\mu_1- \lambda p_1 & - \lambda p_1 & \dots & - \lambda p_1\\
        -\mu_2 & -\mu_2- \lambda p_2 & \dots & - \lambda p_2\\
        \dots & \dots & \dots & \dots \\
        \mu_r- \lambda p_r & \mu_r- \lambda p_r & \dots & -\lambda p_r
        \end{bmatrix},\\
    \mathbf{c_a} &= \begin{bmatrix}
        - \lambda p_1 \\
        - \lambda p_2\\
        \dots  \\
        - \lambda p_r  
        \end{bmatrix} ,
    \mathbf{c_b} = \begin{bmatrix}
        - \lambda p_1 \\
        - \lambda p_2\\
        \dots  \\
        - \lambda p_r  + k \alpha \mu_r
        \end{bmatrix}.
\end{align*}
\end{theorem}
\begin{proof} 
When $k_i=k$, $\alpha_1=\alpha$ and $d=1$, we can suppress the service stage index $j$ and the ODE's from \eqref{eq:diffMultServMultType} reduces to:

\begin{align}\label{eq:diffOneServMultType}
    \dot w_{1} &= \lambda p_1 \brac{1-\sum_{a} w_{a}}- \mu_{1} w_{1}, \nonumber \\ 
    \dot w_{i} &= \lambda p_i \brac{1-\sum_{a} w_{a}}- \mu_{i} \min \brac{w_{i},\max \brac{0,k \alpha-\sum_{l<i}w_{l}}}, 2\le i \le r.
\end{align}
Next we show that $\mathbf{B}$ is non-singular and real parts of its eigenvalues are negative. Same holds for $\mathbf{A}$ which can be proved in a similar, although simpler, way.

Let $\mathbf{Bx} = \mathbf{0}$ with $\mathbf{x} \ne \mathbf{0}$. Then 
\begin{align*}
     \begin{bmatrix}
        \mu_1 x_1\\
        \mu_2 x_2\\
        \dots \\
        \mu_r x_r
    \end{bmatrix} & = - \sum_l x_l
    \begin{bmatrix}
        \lambda p_1\\
        \lambda p_2\\
        \dots \\
        \lambda p_r - \mu_r
    \end{bmatrix}.    
\end{align*}
Since $\mathbf{x} \ne \mathbf{0}$, we have $\sum_l x_l \ne 0$ and
\begin{align*}
    & \sum_l x_l = -\lambda \brac{\sum_l x_l } \brac{\sum_l \frac{p_l}{\mu_l} }+ \sum_l x_l \\
   \text{i.e.,} \quad & \lambda \sum_l \frac{p_l}{\mu_l}  = 0,
\end{align*}
which contradicts positivity of $\lambda$, $\mu_i$'s and $p_i$'s.

Next we prove that the eigenvalues of $\mathbf{B}$ have negative real part. Let $\theta$ be an eigenvalue and $\mathbf{u}$ be a corresponding eigenvector. For $\theta \ne -\mu_j ~\forall j$, we have
\begin{align*}
    & u_j(\theta+ \mu_j) = -\lambda p_j \sum_{l} u_l, \quad  1 \le j \le r-1 \\
    \text{and} \quad & u_r(\theta+ \mu_r) = \brac{-\lambda p_r + \mu_r}  \sum_{l} u_l.
\end{align*}
Since $\theta \ne -\mu_j ~\forall j$ and $\mathbf{u} \ne \mathbf{0}$, we have $\sum_{l} u_l \ne 0$ and

\begin{align*}
     &\sum_l u_l = \brac{-\frac{\mu_r}{\mu_r + \theta}+ \sum_l \frac{- \lambda p_l}{\mu_l + \theta}} \brac{\sum_l u_l}, \\
    \text{i.e.,}~ &
     \sum_l \frac{\lambda p_l}{\mu_l + \theta} = -1 +\frac{\mu_r}{\mu_r + \theta}, \\
    \text{i.e.,}~ &
    \sum_l \frac{\lambda p_l (\mu_l + \Re{(\theta)})}{|\mu_l + \theta|^2} = -1 + \frac{\mu_r (\mu_r + \Re{(\theta)})}{|\mu_r + \theta|^2}.
\end{align*}

The left and right hand sides have different signs unless $\Re{(\theta)}<0$. If $\theta = -\mu_j$ for some $j$, we are done anyway. Now we return to \eqref{eq:diffOneServMultType} and prove global attraction to the unique fixed point under different scenarios.

{\bf Case 1}: $k \alpha \ge 1$ \\ 
In this case, \eqref{eq:diffOneServMultType} becomes
\begin{align*}
     \frac{d\mathbf{w}}{dt}= \mathbf{A} \mathbf{w} - \mathbf{c_a}.
\end{align*}
and the system is globally attractive to the unique fixed point $\mathbf{A}^{-1} \mathbf{c_a}$ as $\mathbf{A}$ is non-singular and all its eigenvalues have negative real part.

{\bf Case 2}: $k \alpha < 1$ \\ 
We start by showing $\langle \mathbf{A}^{-1} \mathbf{c_a}, 1 \rangle < k \alpha$ iff $\langle \mathbf{B}^{-1} \mathbf{c_b}, 1 \rangle < k \alpha$. For if $\mathbf{A} \mathbf{x} = \mathbf{c_a}$ and $\mathbf{B} \mathbf{y} = \mathbf{c_b}$, we have 
\begin{align*}
    \sum_l x_l &= \frac{\sum_l \frac{\lambda p_l}{\mu_l}}{1+\sum_l \frac{\lambda p_l}{\mu_l}} \\
    \text{and}~ 
    \sum_l y_l &= \frac{\sum_l \frac{\lambda p_l}{\mu_l}- k \alpha}{\sum_l \frac{\lambda p_l}{\mu_l}}.
\end{align*}
And 
\begin{align*}
    \frac{\sum_l \frac{\lambda p_l}{\mu_l}}{1+\sum_l\frac{\lambda p_l}{\mu_l}} < k \alpha \iff
    \frac{\sum_l \frac{\lambda p_l}{\mu_l}- k \alpha}{\sum_l \frac{\lambda p_l}{\mu_l}} < k \alpha.
\end{align*}
Next, we observe that the system eventually enters the region $\sum_{i \le r-1} w_{i} < k \alpha$. This is because existence of a lowest index $i_0<r-1$ with $\sum_{i\le i_0} w_{i} \ge k \alpha$ implies $\frac{dw_i}{dt} \ge 0$ for $i \ge i_0$, since the domain of interest is $\sum_{i} w_{ij} \le 1$. This is an unstable system with $w_i$, $i>i_0$ increasing forever and thus it eventually enters the region $\sum_{i \le r-1} w_{i}< k \alpha$.

Let us assume  $\langle \mathbf{A}^{-1} \mathbf{c_a}, 1 \rangle < k \alpha$. If we start the system in the subregion $\sum_{i \le r} w_{i}< k \alpha$, the system evolves in a fashion similar to the case $k \alpha \ge 1$ and converges to $ \mathbf{A}^{-1} \mathbf{c_a}$. When the system is started in the subregion $\sum_{i \le r} w_{i} \ge k \alpha$, the evolution is given by
\begin{align*}
     \frac{d\mathbf{w}}{dt}= \mathbf{B} \mathbf{w} - \mathbf{c_b}.
\end{align*}
We see that $\mathbf{B}$ is non singular and the eigenvalues have negative real part. Hence the system move toward the point $ \mathbf{B}^{-1} \mathbf{c_b}$. However,  $\langle \mathbf{A}^{-1} \mathbf{c_a}, 1 \rangle < k \alpha$ implies  $\langle \mathbf{B}^{-1} \mathbf{c_b}, 1 \rangle < k \alpha$. Hence, the system eventually enters the subregion $\sum_{i \le r} w_{i}< k \alpha$ and converges to $ \mathbf{A}^{-1} \mathbf{c_a}$. For the case $\langle \mathbf{A}^{-1} \mathbf{c_a}, 1 \rangle > k \alpha$, the system converges to $ \mathbf{B}^{-1} \mathbf{c_b}$ and the proof proceeds similarly.
\end{proof}
\newpage



\end{document}